\documentclass[traditabstract]{aa}
\usepackage{graphicx}
\usepackage{stfloats}                                                   
\usepackage{txfonts}
\newcommand{\nature}{Nature}                                   
\newcommand{\asr}{Adv. Space Res.}                          
\newcommand{\gpn}{Geophys. Norveg.}                      

\begin{document}
\title{Pseudo 2D chemical model of hot-Jupiter atmospheres:\\
application to HD~209458b and HD~189733b}
\titlerunning{Pseudo 2D chemical model of hot-Jupiter atmospheres}
\authorrunning{Ag\'undez et al.}

\author{Marcelino Ag\'undez\inst{1,2}\thanks{Current address: Centro de Astrobiolog\'ia INTA-CSIC, Carretera de Ajalvir, km.4, ES-28850 Madrid, Spain}, Vivien Parmentier\inst{3}, Olivia Venot\inst{4}, Franck Hersant\inst{1,2}, and Franck Selsis\inst{1,2}}

\institute{
Univ. Bordeaux, LAB, UMR 5804,
F-33270, Floirac, France \and 
CNRS, LAB, UMR 5804, F-33270, Floirac, France \and 
Universit\'e de Nice-Sophia Antipolis, Observatoire de la C\^ote d’Azur, CNRS UMR 6202, BP 4229, 06304 Nice Cedex 4, France \and
Instituut voor Sterrenkunde, Katholieke Universiteit Leuven, Celestijnenlaan 200D, 3001 Leuven, Belgium
}

\date{Received; accepted}


\abstract
{The high temperature contrast between the day and night sides of hot-Jupiter atmospheres may result in strong variations of the chemical composition with longitude if the atmosphere were at chemical equilibrium. On the other hand, the vigorous dynamics predicted in these atmospheres, with a strong equatorial jet, would tend to supress such longitudinal variations. To address this subject we have developed a pseudo two-dimensional model of a planetary atmosphere, which takes into account thermochemical kinetics, photochemistry, vertical mixing, and horizontal transport, the latter being modeled as a uniform zonal wind. We have applied the model to the atmospheres of the hot Jupiters HD~209458b and HD~189733b. The adopted eddy diffusion coefficients were calculated by following the behavior of passive tracers in three-dimensional general circulation models, which results in much lower eddy values than in previous estimates. We find that the distribution of molecules with altitude and longitude in the atmospheres of these two hot Jupiters is complex because of the interplay of the various physical and chemical processes at work. Much of the distribution of molecules is driven by the strong zonal wind and the limited extent of vertical transport, resulting in an important homogenization of the chemical composition with longitude. The homogenization is more marked in planets lacking a thermal inversion such as HD~189733b than in planets with a strong stratosphere such as HD~209458b. In general, molecular abundances are quenched horizontally to values typical of the hottest dayside regions, and thus the composition in the cooler nightside regions is highly contaminated by that of warmer dayside regions. As a consequence, the abundance of methane remains low, even below the predictions of previous one-dimensional models, which probably is in conflict with the high CH$_4$ content inferred from observations of the dayside of HD~209458b. Another consequence of the important longitudinal homogenization of the abundances is that the variability of the chemical composition has little effect on the way the emission spectrum is modified with phase and on the changes in the transmission spectrum from the transit ingress to the egress. These variations in the spectra are mainly due to changes in the temperature, rather than in the composition, between the different sides of the planet.}
{}
{}
{}
{}

\keywords{planets and satellites: atmospheres -- planets and satellites: composition -- planets and satellites: individual (HD~209458b) -- planets and satellites: individual (HD~189733b)}

\maketitle

\section{Introduction}

Strongly irradiated by their close host star, hot Jupiters reside in extreme environments and represent a class of planets without analogue in our solar system. This type of exoplanets, the first to be discovered around main-sequence stars, remains the best available to study through observations and challenge a variety of models in the area of planetary science (see reviews on the subject by \cite{bar2010} 2010; \cite{sea2010} 2010; \cite{bur2011} 2011; \cite{sho2011} 2011). In recent years, multiwavelength observations of transiting hot Jupiters have allowed scientists to put constraints on the physical and chemical state of their atmospheres. Among these hot Jupiters, the best characterized are probably HD~209458b and HD~189733b, which belong to some of the brightest and closest transiting systems, and for which primary transit, secondary eclipse, and phase curve measurements have been used to probe, though often with controversial interpretations, various characteristics of their atmospheres, such as the thermal structure (\cite{dem2005} 2005, 2006; \cite{knu2008} 2008; \cite{cha2008} 2008), winds and day-night heat redistribution (\cite{knu2007} 2007, 2012; \cite{cow2007} 2007; \cite{sne2010} 2010), and mixing ratios of some of the main molecular constituents (\cite{tin2007} 2007; \cite{swa2008} 2008, 2009a, 2009b, 2010; \cite{gri2008} 2008; \cite{sin2009} 2009; \cite{des2009} 2009; \cite{bea2010} 2010; \cite{gib2011} 2011; \cite{wal2012} 2012; \cite{lee2012} 2012; \cite{rod2013} 2013; \cite{dek2013} 2013).

The ability of observations at infrared and visible wavelengths to characterize the physical and chemical state of exoplanet atmospheres has motivated the development of various types of theoretical models. On the one hand, there are those aiming at investigating the physical structure of hot-Jupiter atmospheres, either one-dimensional radiative models (\cite{iro2005} 2005; \cite{for2008} 2008; \cite{par2014} 2014) or three-dimensional general circulation models (\cite{cho2008} 2008; \cite{sho2009} 2009; \cite{hen2011a} 2011a,b; \cite{dob2012} 2012; \cite{rau2013} 2013; \cite{par2013} 2013). These models have shown how fascinating the climates of hot Jupiters are, with atmospheric temperatures usually in excess of 1000 K, and helped to understand some global observed trends. Some hot Jupiters are found to display a strong thermal inversion in the dayside while others do not (e.g. \cite{for2008} 2008). Strong winds with velocities of a few km s$^{-1}$ develop and transport the heat from the dayside to the nightside, reducing the temperature contrast between the two hemispheres. The circulation pattern in these planets is characterized by an equatorial superrotating eastward jet. On the other hand, the chemical composition of hot Jupiters has been investigated by one-dimensional models, which currently account for thermochemical kinetics, vertical mixing, and photochemistry (\cite{zah2009} 2009; \cite{lin2010} 2010, 2011; \cite{mos2011} 2011, 2013; \cite{kop2012} 2012; \cite{ven2012} 2012, 2014; \cite{agu2014} 2014). These models have revealed the existence of three different chemical regimes in the vertical direction. A first one at the bottom of the atmosphere, where the high temperatures and pressures ensure a chemical equilibrium composition. A second one located above this, where the transport of material between deep regions and higher layers occurs faster than chemical kinetics so that abundances are quenched at the chemical equilibrium values of the quench level. And a third one located in the upper atmosphere, where the exposure to ultraviolet (UV) stellar radiation drives photochemistry. The exact boundaries between these three zones depend on the physical conditions of the atmosphere and on each species.

In addition to the retrieval of average atmospheric quantities from observations, there is a growing interest in the physical and chemical differences that may exist between different longitudes and latitudes in hot-Jupiter atmospheres, and in the possibility of probing these gradients through observations. Indeed, important temperature contrasts between different planetary sides of hot Jupiters, noticeably between day and night sides, have been predicted (\cite{sho2002} 2002), observed for a dozen hot Jupiters (see \cite{knu2007} 2007 for the first one), qualitatively understood (\cite{cow2011} 2011; \cite{per2013} 2013), and confirmed by three-dimensional general circulation models (e.g. \cite{per2012} 2012). These temperature gradients, together with the fact that photochemistry switches on and off in the day and night sides, are at the origin of a potential chemical differentiation in the atmosphere along the horizontal dimension, especially along longitude. On the other hand, strong eastward jets with speeds of a few km s$^{-1}$ are believed to dominate the atmospheric circulation in the equatorial regions, as predicted by \cite{sho2002} (2002), theorized in \cite{shopol2011} (2011), potentially observed by \cite{sne2010} (2010), and confirmed by almost all general circulation models of hot Jupiters. These strong horizontal winds are an important potential source of homogenization of the chemical composition between locations with different temperatures and UV illumination. The existence of winds and horizontal gradients in the temperature and chemical composition of hot-Jupiter atmospheres has mainly been considered from a theoretical point of view, although some of these effects can be studied through phase curve observations (\cite{for2006} 2006; \cite{cow2008} 2008; \cite{maj2012} 2012; \cite{dew2012} 2012), monitoring of the transit ingress and egress (\cite{for2010} 2010), and Doppler shifts of spectral lines during the primary transit (\cite{sne2010} 2010; \cite{mil2012} 2012; \cite{sho2013} 2013).

The existence of horizontal chemical gradients has been addressed in the frame of a series of one-dimensional models in the vertical direction at different longitudes (e.g. \cite{mos2011} 2011). An attempt to understand the interplay between circulation dynamics and chemistry was undertaken by \cite{coo2006} (2006), who coupled a three-dimensional general circulation model of HD 209458b to a simple chemical kinetics scheme dealing with the interconversion between CO and CH$_4$. These authors found that, even in the presence of strong temperature gradients, the mixing ratios of CO and CH$_4$ are homogenized throughout the planet's atmosphere in the 1 bar to 1 mbar pressure regime. In our team, we have recently adopted a different approach in which we coupled a robust chemical kinetics scheme to a simplified dynamical model of HD~209458b's atmosphere (\cite{agu2012} 2012). In this approach the atmosphere was assumed to rotate as a solid body, mimicking a uniform zonal wind, while vertical mixing and photochemistry were neglected. We found that the zonal wind acts as a powerful disequilibrium process that tends to homogenize the chemical composition, bringing molecular abundances at the limb and nightside regions close to chemical equilibrium values characteristic of the dayside. Here we present an improved model that simultaneously takes into account thermochemical kinetics, photochemistry, vertical mixing, and horizontal transport in the form of a uniform zonal wind. We apply our model to study the interplay between atmospheric dynamics and chemical processes, and the distribution of the main atmospheric constituents in the atmosphere of the hot Jupiters HD~209458b and HD~189733b.

\section{Model} \label{sec:model}

We modeled the atmospheres of HD~209458b and HD~189733b, for which we adopted the parameters derived by \cite{sou2010} (2010). For the system of HD~209458 we took a stellar radius of 1.162 $R_{\odot}$, a planetary radius and mass of 1.38 $R_J$ and 0.714 $M_J$ (where $R_J$ and $M_J$ stand for Jupiter radius and mass), and an orbital distance of 0.04747 au. For the system of HD~189733 the adopted parameters are a stellar radius of 0.752 $R_{\odot}$, a planetary radius and mass of 1.151 $R_J$ and 1.150 $M_J$, and a planet-to-star distance of 0.03142 au.

The atmosphere model is based on some of the outcomes of three-dimensional general circulation models (GCMs) developed for HD~209458b and HD~189733b (\cite{sho2009} 2009; \cite{par2013} 2013, in preparation), which indicate that circulation dynamics is dominated by a broad eastward equatorial jet. On the assumption that the eastward jet dominates the circulation pattern, it seems well justified to model the atmosphere as a vertical column that rotates along the equator, which mimicks a uniform zonal wind. The main shortcoming of this approach is that it reduces the whole circulation dynamics to a uniform zonal wind, although it has the clear advantage over more traditional one-dimensional models in the vertical direction of simultaneously taking into account the mixing and transport of material in the vertical and horizontal directions.

\subsection{Pseudo two-dimensional chemical model}

In one-dimensional models of planetary atmospheres, the distribution of each species in the vertical direction is governed by the coupled continuity-transport equation
\begin{equation}
\frac{\partial f_i}{\partial t} = \frac{P_i}{n} - f_i L_i - \frac{1}{n r^2} \frac{\partial (r^2 \phi_i)}{\partial r}, \label{eq:continuity}
\end{equation}
where $f_i$ is the mixing ratio of species $i$, $t$ the time, $n$ the total number density of particles, $r$ the radial distance to the center of the planet, $P_i$ and $L_i$ the rates of production and loss, respectively, of species $i$, and $\phi_i$ the vertical transport flux of particles of species $i$ (positive upward and negative downward). The first two terms on the right side of Eq.~(\ref{eq:continuity}) account for the formation and destruction of species $i$ by chemical and photochemical processes, while the third term accounts for the vertical transport in a spherical atmosphere. In this way, thermochemical kinetics, photochemistry, and vertical mixing can be taken into account through Eq.~(\ref{eq:continuity}). The transport flux can be described by eddy and molecular diffusion as
\begin{equation}
\phi_i = - K_{zz} n \frac{\partial f_i}{\partial z} - D_i n \Big( \frac{\partial f_i}{\partial z} + \frac{f_i}{H_i} - \frac{f_i}{H_0} + \frac{\alpha_i}{T} \frac{d T}{ d z} f_i \Big), \label{eq:flux}
\end{equation}
where $z$ is the altitude in the atmosphere with respect to some reference level (typically set at a pressure of 1 bar), $T$ is the gas kinetic temperature, $K_{zz}$ is the eddy diffusion coefficient, $D_i$ is the coefficient of molecular diffusion of species $i$, $H_i$ is the scale height of species $i$, $H_0$ is the mean scale height of the atmosphere, and $\alpha_i$ is the thermal diffusion factor of species $i$. More details on Eqs.~(\ref{eq:continuity}) and (\ref{eq:flux}) can be found, for instance, in \cite{bau1973} (1973) and \cite{yun1999} (1999). The coefficient of molecular diffusion $D_i$ is estimated from the kinetic theory of gases (see \cite{rei1988} 1988), while the factor of thermal diffusion $\alpha_i$ is set to $-0.25$ for the light species H, H$_2$, and He (\cite{bau1973} 1973), and to 0 for the rest of species. The eddy diffusion coefficient $K_{zz}$ is a rather empirical formalism to take into account advective and turbulent mixing processes in the vertical direction, and is discussed in more detail in section~\ref{subsec:gcm}.

To compute the abundances of the different species as a function of altitude, the atmosphere is divided into a certain number of layers and the continuous variables in Eqs.~(\ref{eq:continuity}) and (\ref{eq:flux}) are discretized as a function of altitude. After the discretization, Eq.~(\ref{eq:continuity}) reads
\begin{equation}
\frac{\partial f_i^j}{\partial t} = \frac{P_i^j}{n^j} - f_i^j L_i^j - \frac{\big(r^{j+1/2}\big)^2 \phi_i^{j+1/2} - \big(r^{j-1/2}\big)^2 \phi_i^{j-1/2}}{n^j \big(r^j\big)^2 \big(z^{j+1/2} - z^{j-1/2}\big)}, \label{eq:continuity-discrete}
\end{equation}
where the superscript $j$ refers to the $j^{\rm th}$ layer, while $j+1/2$ and $j-1/2$ refer to its upper and lower boundaries, respectively, so that layers are ordered from bottom to top. The transport fluxes of species $i$ at the upper and lower boundaries of layer $j$, $\phi_i^{j+1/2}$ and $\phi_i^{j-1/2}$, are then given by
\begin{eqnarray}
\phi_i^{j \pm 1/2} = - K_{zz}^{j \pm 1/2} n^{j \pm 1/2} \frac{\partial f_i}{\partial z} \Big|_{j \pm 1/2} - D_i^{j \pm 1/2} n^{j \pm 1/2} \Bigg[ \frac{\partial f_i}{\partial z} \Big|_{j \pm 1/2} \nonumber \\
+ \bigg( \frac{f_i^{j \pm 1/2}}{H_i^{j \pm 1/2}} - \frac{f_i^{j \pm 1/2}}{H_0^{j \pm 1/2}} + \frac{\alpha_i}{T^{j \pm 1/2}} \frac{d T}{d z} \Big|_{j \pm 1/2} f_i^{j \pm 1/2} \bigg) \Bigg], \label{eq:flux-discrete}
\end{eqnarray}
where the variables evaluated at $j+1/2$ and $j-1/2$ boundaries are approximated as the arithmetic mean of the values at layers $j$ and $j+1$ and at layers $j-1$ and $j$, respectively. We assume that there is neither gain nor loss of material in the atmosphere, and thus the transport fluxes at the bottom and top boundaries of the atmosphere are set to zero.

The rates of production and loss of each species in Eqs.~(\ref{eq:continuity}) and (\ref{eq:continuity-discrete}) are given by chemical and photochemical processes. Thermochemical kinetics is taken into account with a chemical network, which consists of 104 neutral species composed of C, H, N, and O linked by 1918 chemical reactions, that has been validated in the area of combustion chemistry by numerous experiments over the 300-2500 K temperature range and the 0.01-100 bar pressure regime, and has been found suitable to model the atmospheres of hot Jupiters. Most reactions are reversed with their rate constants fulfilling detailed balance to ensure that, in the absence of disequilibrium processes such as photochemistry or mixing, thermochemical equilibrium is achieved at sufficiently long times. The reaction scheme is described in \cite{ven2012} (2012), with some minor modifications given in \cite{agu2012} (2012). As photochemical processes we consider photodissociations, whose rates depend on the incident UV flux and the relevant cross sections. The incident UV flux is calculated by solving the radiative transfer in the vertical direction for a given zenith angle, where the spherical geometry of layers is taken into account when computing the path length along each of them. Absorption and Rayleigh scattering, the latter being treated through a two-ray iterative algorithm (\cite{isa1977} 1977), both contribute to the attenuation of UV light throughout the atmosphere. Absorption and photodissociation cross-sections are described in detail in \cite{ven2012} (2012). Rayleigh-scattering cross-sections are calculated for the most abundant species from their polarizability (see e.g. \cite{tar1969} 1969). As UV spectrum for the host star HD~209458, we adopt the spectrum of the Sun (mean between minimum and maximum activity from \cite{thu2004} 2004) below 168 nm and a Kurucz synthetic spectrum\footnote{See \texttt{http://kurucz.harvard.edu/stars/hd209458}} at longer wavelengths. For HD~189733b, below 335 nm we adopt a UV spectrum of $\epsilon$ Eridani based on the CAB X-exoplanets archive (\cite{san2011} 2011) and observations with FUSE and HST (see details in \cite{ven2012} 2012), and a Kurucz synthetic spectrum\footnote{See \texttt{http://kurucz.harvard.edu/stars/hd189733}} above 335 nm. 

In one-dimensional vertical models of planetary atmospheres, the system of differential equations given by Eq.~(\ref{eq:continuity-discrete}), with as many equations as the number of layers times the number of species, is integrated as a function of time, starting from some initial composition, usually given by thermochemical equilibrium, until a steady state is reached. During the evolution, the physical conditions of the vertical atmosphere column remain static. In the pseudo two-dimensional approach adopted here, we consider that the vertical atmosphere column rotates around the planet's equator, and thus the system of differential equations is integrated as a function of time with physical conditions varying with time, according to the periodic changes experienced during this travel. A vertical atmosphere column rotating around the equator mimics a uniform zonal wind, which is an idealization of the equatorial superrotating jet structure found by three-dimensional GCMs for hot-Jupiter atmospheres. This approach may be seen as a pseudo two-dimensional model in which the second dimension, which corresponds to the longitude (the first one being the altitude), is in fact treated as a time dependence in the frame of an atmosphere column rotating around the equator. 

To build the pseudo two-dimensional chemical models of HD~209458b and HD~189733b the vertical atmosphere column is divided into 100-200 layers spanning the pressure range 500-10$^{-8}$ bar. The evolution of the vertical atmosphere column starts at the substellar point with an initial composition given by either thermochemical equilibrium or a one-dimensional vertical model, the latter usually resulting in shorter integration times before a periodic state is reached. The convenience of starting with the composition of the hottest substellar regions is discussed in \cite{agu2012} (2012). Thermochemical equilibrium calculations were carried out using a code that minimizes the Gibbs energy based on the algorithm of \cite{gor1994} (1994) and the thermochemical data described in \cite{ven2012} (2012) for the 102 species included. A solar elemental composition (\cite{asp2009} 2009) was adopted for the atmospheres of both HD~209458b and HD~189733b. The planetary sphere was then discretized into a certain number of longitudes (typically 100) and the system of differential equations given by Eq.~(\ref{eq:continuity-discrete}) was integrated as the atmosphere column moves from one longitude to the next, at a constant angular velocity. To speed up the numerical calculations, the physical variables that vary with longitude (in our case these are the vertical structures of temperature and incident UV flux) were discretized as a function of longitude, that is, they were assumed to remain constant within each discretized longitude interval. As long as there are important longitudinal temperature gradients, the atmospheric scale height also varies with longitude, so that the atmosphere expands or shrinks depending on whether it gets warmer or cooler. To incorporate this effect, which may have important consequences for transit spectra, the vertical atmosphere column was enlarged or compressed (the radius at the base of the atmosphere remaining fixed) to fulfill hydrostatic equilibrium at any longitude. The variation of the incident UV flux with longitude was taken into account through the zenith angle. At the limbs we considered a zenith angle slightly different from a right angle because of the finite apparent size of the star and because of atmospheric refraction, for which we adopted a refraction angle of half a degree as in the case of visible light at Earth.

The nonlinear system of first-order ordinary differential equations given by Eq.~(\ref{eq:continuity-discrete}) was integrated as a function of time using a backward differentiation formula implicit method for stiff problems implemented in the Fortran solver DLSODES within the ODEPACK package\footnote{See \texttt{http://computation.llnl.gov/casc/odepack}} (\cite{hin1983} 1983; \cite{rad1993} 1993). The evolution of the vertical atmosphere column was followed during several rotation cycles until the abundances of the main atmospheric constituents achieved a periodic behavior, which for HD~209458b and HD~189733b, occurs after some tens or hundreds of rotation periods.

\subsection{Atmospheric dynamics and temperature (GCMs)} \label{subsec:gcm}

\begin{figure}
\centering
\includegraphics[angle=0,width=\columnwidth]{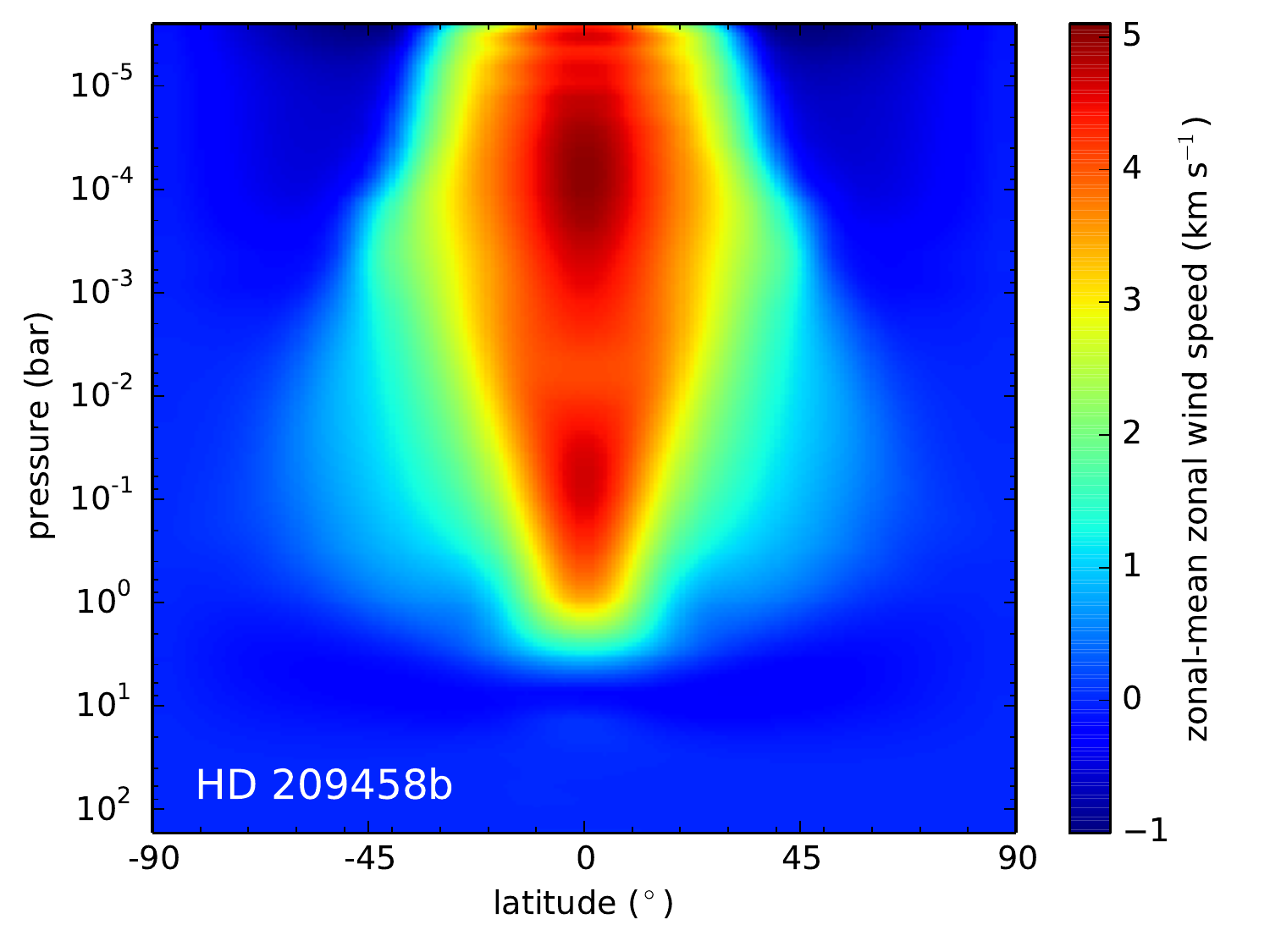}
\caption{Zonal-mean zonal wind speed (positive is eastward and negative westward) as a function of latitude and pressure, as calculated with a GCM simulation of HD~209458b (\cite{par2013} 2013). Note the superrotating wind above the 1 bar pressure level in the equatorial region ($\pm$20$^\circ$ in latitude).} \label{fig:uzonal-hd209458b}
\end{figure}

The pseudo two-dimensional chemical model needs some key input data related to the zonal wind speed, thermal structure, and strength of vertical mixing. These data are calculated with the three-dimensional general circulation model SPARC/MITgcm developed by \cite{sho2009} (2009), in which dynamics and radiative transfer are coupled. The data used here for HD~209458b are based on the simulations by \cite{par2013} (2013), while those for HD~189733b are based on calculations by Parmentier et al. (in preparation), both of which cover a pressure range from about 200 bar to 2 $\mu$bar. These GCM simulations provide a wealth of detailed information regarding the physical structure of the atmosphere, although they remain limited with respect to the chemical structure as long as the composition is assumed to be given by local chemical equilibrium.

\begin{figure}
\centering
\includegraphics[angle=0,width=\columnwidth]{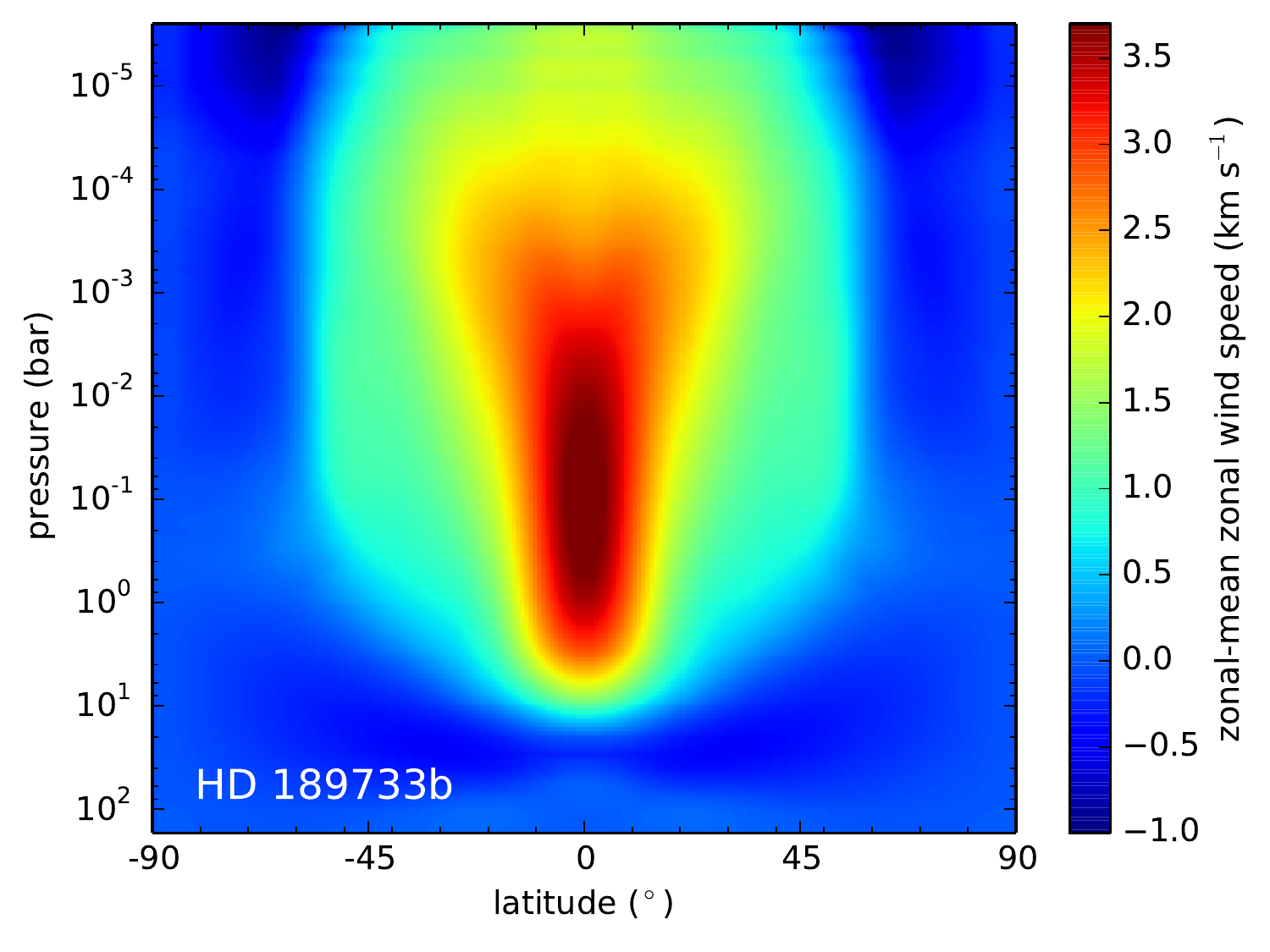}
\caption{Zonal-mean zonal wind speed as a function of latitude and pressure calculated with a GCM simulation of HD~189733b (Parmentier et al. in preparation). A strong superrotating equatorial jet is clearly present in the 10-10$^{-3}$ bar pressure range.} \label{fig:uzonal-hd189733b}
\end{figure}

\subsubsection{Wind structure} \label{subsubsec:wind}

Circulation dynamics in the atmospheres of hot Jupiters is dominated by a fast eastward (or superrotating) jet stream at the equator. This superrotating jet was first predicted by \cite{sho2002} (2002), has been found to emerge from almost all GCM simulations of hot Jupiters (\cite{coo2005} 2005; \cite{sho2008} 2008, 2009, 2013; \cite{dob2008} 2008; \cite{rau2010} 2010, 2012a,b; \cite{per2010} 2010, 2012; \cite{hen2011a} 2011a,b; \cite{lew2010} 2010; \cite{kat2013} 2013; \cite{par2013} 2013), and is also understood theoretically (\cite{shopol2011} 2011). A shift of the hottest point of the planet eastward from the substellar point has been directly observed in several exoplanets (\cite{knu2007} 2007, 2009a, 2012; \cite{cro2010} 2010) and interpreted as a direct consequence of this jet. The superrotating jet in HD~209458b and HD~189733b spans over all longitudes and has a well-defined location in latitude (around $\pm$20$^\circ$) and pressure (between 1-10 bar and 10$^{-6}$-10$^{-3}$ bar), as illustrated in Figs.~\ref{fig:uzonal-hd209458b} and \ref{fig:uzonal-hd189733b}, where the zonally averaged zonal wind speed is depicted as a function of latitude and pressure for each planet. To derive a mean speed of the jet for our pseudo two-dimensional chemical model, we averaged the zonal wind speed longitudinally over the whole planet, latitudinally over $\pm$20$^\circ$, and vertically between 1 and 10$^{-6}$ bar, the latter corresponding to the top of the atmosphere in the GCM simulations. We find mean zonal wind speeds of 3.85 km s$^{-1}$ for HD~209458b and 2.43 km s$^{-1}$ for HD~189733b, in both cases in the eastward direction. These values were adopted in the pseudo two-dimensional chemical model as the speed of the zonal wind at the equator and 1 bar pressure level, thus setting the angular velocity of the rotating vertical atmosphere column (i.e., its rotation period).

At high latitudes, above 50$^\circ$, the circulation is no longer dominated by the superrotating jet; the zonal-mean zonal wind is westward and the flow exhibits a complex structure, with westward and eastward winds, and a substantial day-to-night flow over the poles (\cite{sho2009} 2009). At low latitudes, the zonal-mean zonal wind is eastward over most of the vertical structure (above the 1-10 bar pressure level), although the shape of the superrotating jet changes gradually with altitude, from a well-defined banded flow with little longitudinal variability of the jet speed in the deep atmosphere to a less banded flow with important longitudinal variations of the wind speed in the upper levels (\cite{sho2009} 2009). It is also worth noting that according to \cite{sho2013} (2013), the circulation regime in the atmospheres of hot Jupiters changes from a superrotating one to a high-altitude day-to-night flow when the radiative or the frictional time scales become short, as occurs at low pressures under intense insolation or strong drag forces. In this regard, we note that the GCM simulations by \cite{par2013} (2013, in preparation) used here are based on a drag-free case, and are thus the most favorable for the presence of a strong equatorial jet. That is, we would have found a somewhat slower equatorial jet if drag forces were included in the GCM simulations (\cite{rau2012a} 2012a, 2013; \cite{sho2013} 2013).

\begin{figure}
\centering
\includegraphics[angle=0,width=\columnwidth]{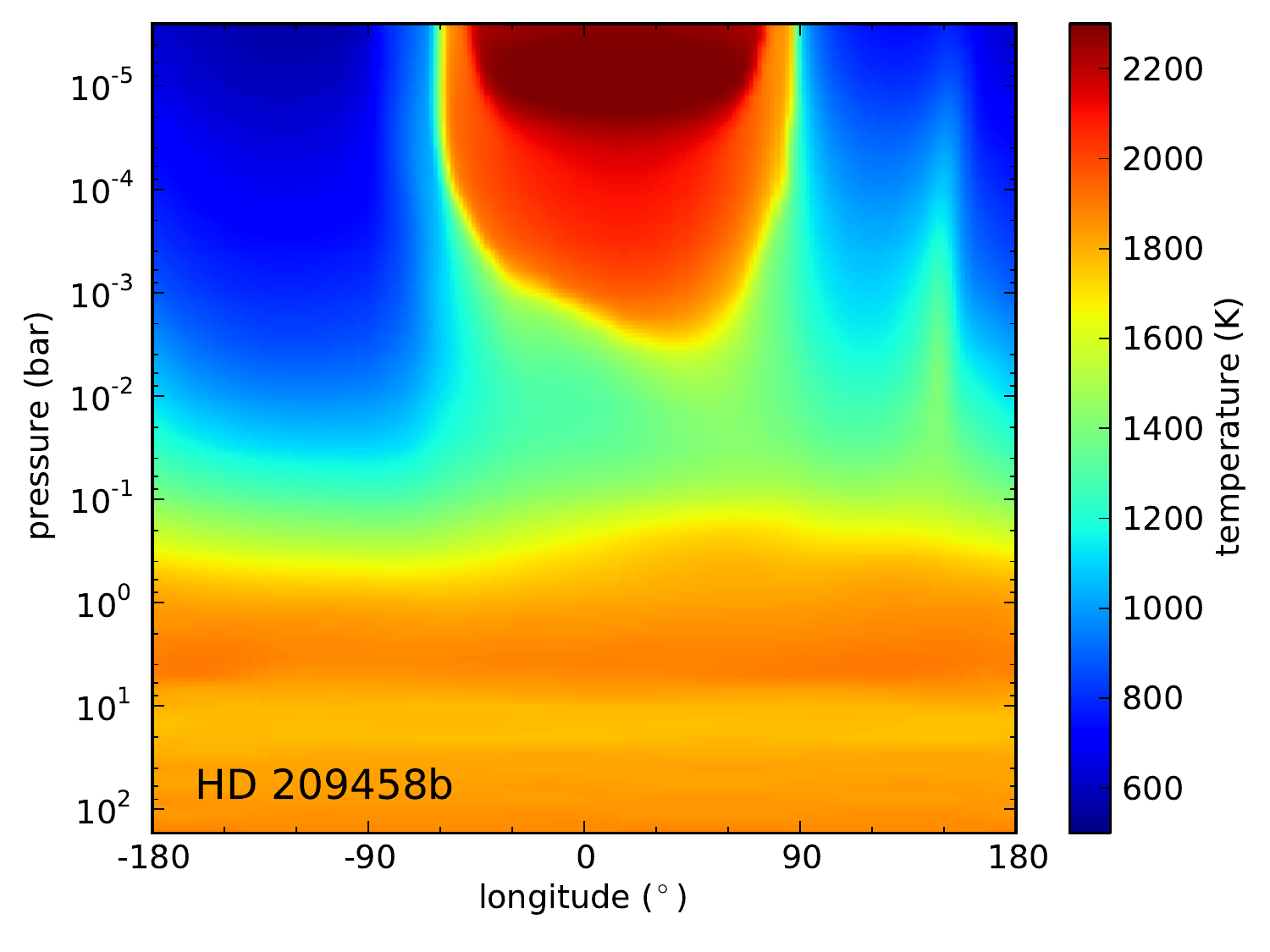}
\caption{Temperature structure averaged latitudinally over $\pm$20$^\circ$ around the equator of HD~209458b, as calculated with a GCM simulation (\cite{par2013} 2013). Note the extremely hot dayside stratosphere above the 1 mbar pressure level.} \label{fig:tk-hd209458b}
\end{figure}

In view of the discussion above, our pseudo two-dimensional chemical model based on a uniform zonal wind probably is a good approximation for the equatorial region ($\pm$20$^\circ$) in the 1 bar to 1 mbar pressure regime, and may still provide a reasonable description of upper equatorial layers, where an eastward jet is still present although with a less uniform structure. In the polar regions our formalism may not be adequate since the circulation regime is more complex, and thus the interplay between dynamics and chemistry may lead to a very different distribution of the chemical composition from that predicted by our model. It is interesting to note that low latitudes contribute more to the projected area of the planet's disk than polar regions, and thus planetary emission is to a large extent dominated by the equatorial regions modeled here. The same is not true for transmission spectra however, where low and high latitudes are both important.

\subsubsection{Temperature structure} \label{subsubsec:tk}

Among the dozen hot Jupiters for which we have good observational constraints on their atmospheric properties (\cite{sea2010} 2010), half of them are believed to have a strong thermal inversion at low pressure in the dayside, while the other half are thought to lack such an inversion. The presence of a stratosphere in hot Jupiters is commonly attributed to the survival in the gas phase of the strong absorbers at visible wavelengths TiO and VO (\cite{for2008} 2008; \cite{sho2009} 2009; \cite{par2013} 2013). In this theoretical framework, planets that are warm enough to have an appreciable opacity due to TiO and VO (pM class planets) host a stratosphere, while those that are cooler (pL class planets) do not develop a temperature inversion in their atmospheres (\cite{for2008} 2008). This, not yet firmly established however because no unambigous detection of TiO has been obtained (\cite{des2008} 2008), the nature of the absorbers that cause temperature inversions in hot Jupiters is still debated. For example, photochemical products of some undetermined nature or arising from the photochemical destruction of H$_2$S have also been postulated as possible absorbers responsible for these stratospheres (\cite{bur2008} 2008; \cite{zah2009} 2009). Moreover, not all planets fit into this pM/pL scheme, and other parameters such as the atmospheric elemental C/O abundance ratio (\cite{mad2012} 2012) or the stellar activity (\cite{knu2010} 2010) might control whether there are stratospheres in hot Jupiters.

\begin{figure}
\centering
\includegraphics[angle=0,width=\columnwidth]{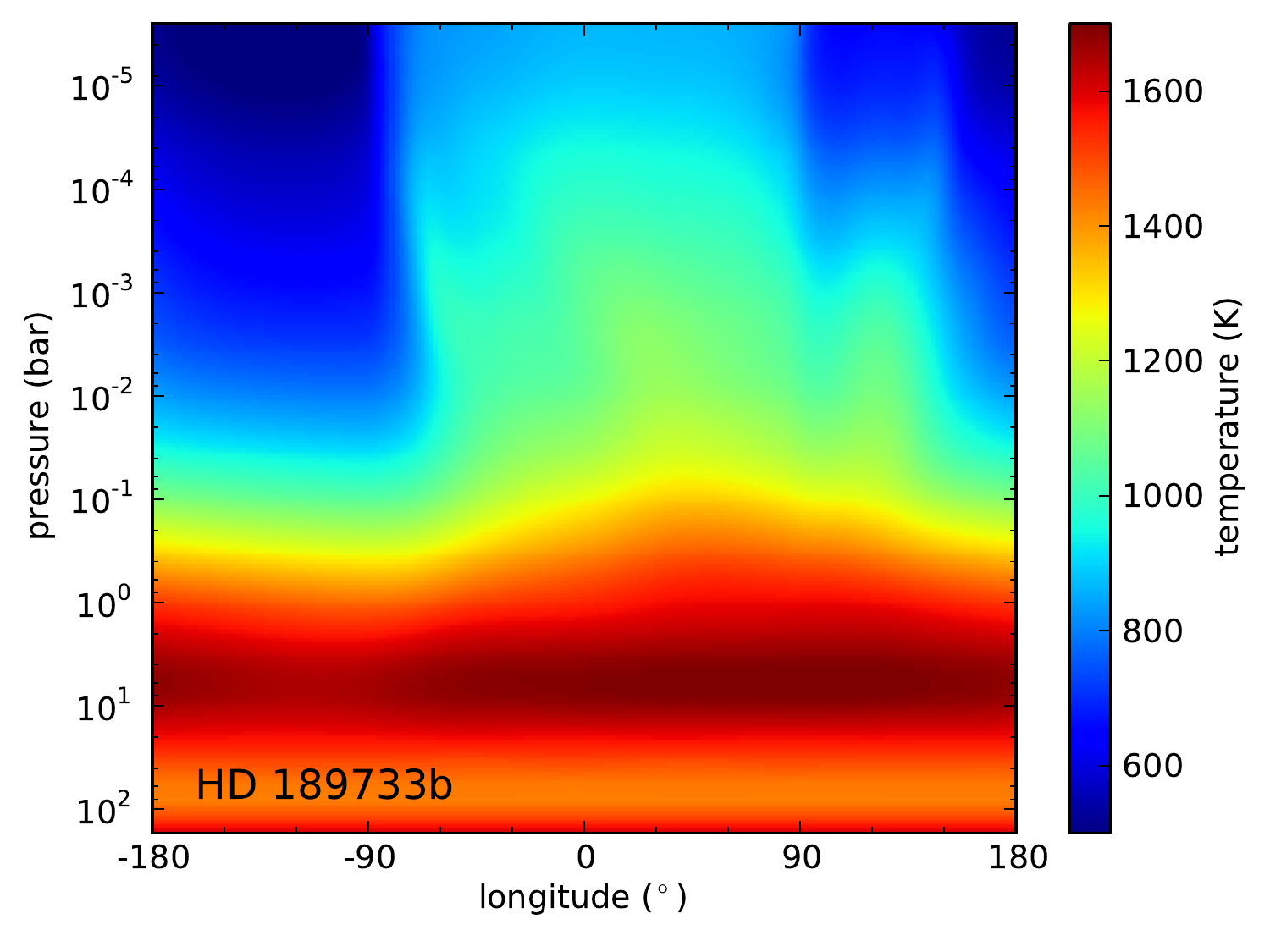}
\caption{Temperature structure averaged latitudinally over $\pm$20$^\circ$ around the equator of HD~189733b, as calculated with a GCM simulation (Parmentier et al. in preparation).} \label{fig:tk-hd189733b}
\end{figure}

HD~209458b and HD~189733b are good examples of these two types of hot Jupiters, the former hosting a strong thermal inversion in the dayside, while the latter does not. The temperature resulting from the GCM simulations and averaged latitudinally over an equatorial band $\pm$20$^\circ$ in latitude is shown as a function of longitude and pressure in Fig.~\ref{fig:tk-hd209458b} for HD~209458b and in Fig.~\ref{fig:tk-hd189733b} for HD~189733b. This equatorial band of $\pm$20$^\circ$ in latitude corresponds to the region where the equatorial jet is present in the GCM simulations (see Figs.~\ref{fig:uzonal-hd209458b} and \ref{fig:uzonal-hd189733b}). For the pseudo two-dimensional chemical model, which focuses on the equatorial region where the eastward jet develops, we adopted the temperature distribution shown in Figs.~\ref{fig:tk-hd209458b} and \ref{fig:tk-hd189733b}, assuming an isothermal atmosphere at pressures lower than 2 $\mu$bar.

In our previous study (\cite{agu2012} 2012), the temperature structure of HD~209458b's atmosphere was calculated with a one-dimensional time-dependent radiative model and resulted in an atmosphere without a strong temperature inversion. Here, the temperature structure calculated for the atmospheres of HD~209458b and HD~189733b comes from GCM simulations, which result in a strong temperature inversion for the former planet and an atmosphere without stratosphere for the latter one. This permits us to explore the chemistry of hot Jupiters with and without a stratosphere. It is also worth noting that in the case of HD~209458b, there is evidence of a dayside temperature inversion from observations of the planetary emission spectrum at infrared wavelengths (\cite{knu2008} 2008).

\subsubsection{Vertical eddy diffusion coefficient} \label{subsubsec:eddy}

\begin{figure}
\centering
\includegraphics[angle=0,width=\columnwidth]{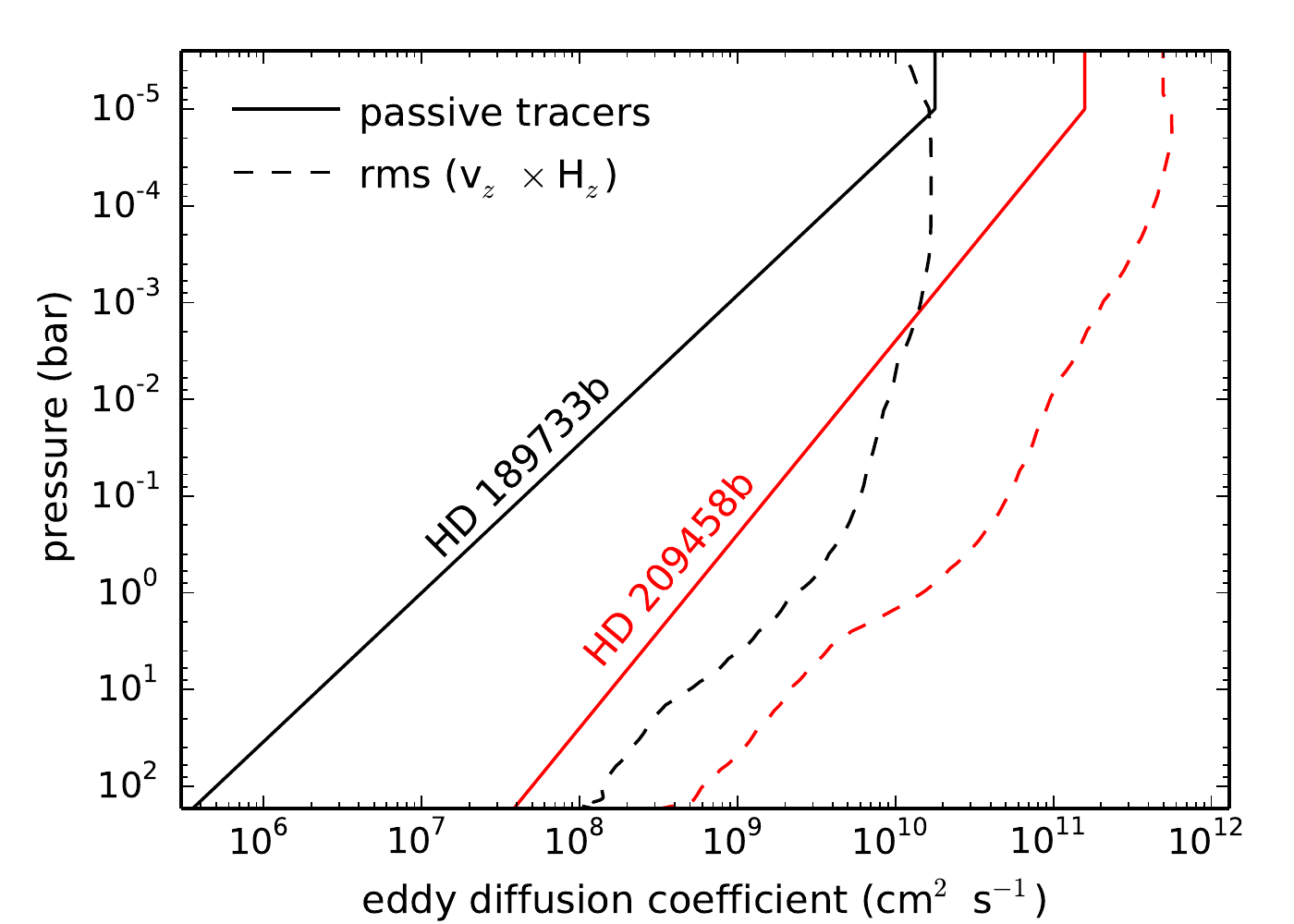}
\caption{Vertical eddy diffusion coefficient profiles for HD~209458b and HD~189733b, as calculated by following the behavior of passive tracers (solid lines; \cite{par2013} 2013, in preparation), and as given by previous estimates based on the rms of the vertical velocity times the vertical scale height (dashed lines; \cite{mos2011} 2011).} \label{fig:eddy}
\end{figure}

Another important outcome of GCM simulations is the quantification of the strength with which material is transported in the vertical direction in the atmosphere. Although this mixing is not diffusive in a rigorous sense, once averaged over the whole planet, it can be well represented by an effective eddy diffusion coefficient that varies with pressure (\cite{par2013} 2013). This variable enters directly as input into one-dimensional and pseudo two-dimensional chemical models of planetary atmospheres such as ours. The eddy diffusion coefficient is commonly estimated in the literature as the root mean square of the vertical velocity times the vertical scale height (\cite{lin2010} 2010; \cite{mos2011} 2011). Recently, \cite{par2013} (2013) have used a more rigorous approach to estimate an effective eddy diffusion coefficient in HD~209458b by following the behavior of passive tracers in a GCM. These authors have shown that vertical mixing in hot-Jupiter atmospheres is driven by large-scale circulation patterns. There are large regions with ascending motions and large regions with descending motions, some of them contributing more to the global mixing than others. It has been also shown that a diffusion coefficient is a good representation of the vertical mixing that takes place in the three-dimensional model of the atmosphere. The resulting values for HD~209458b are 10-100 times lower than those obtained with the previous method (see Fig.~\ref{fig:eddy}), and are used here. The vertical profile of the eddy diffusion coefficient for HD~209458b can be approximated by the expression $K_{zz}$ (cm$^2$ s$^{-1}$) = 5 $\times$ 10$^8$ $p^{-0.5}$, where the pressure $p$ is expressed in bar (see \cite{par2013} 2013). In the case of HD~189733b we used the expression $K_{zz}$ (cm$^2$ s$^{-1}$) = 10$^7$ $p^{-0.65}$, where the pressure $p$ is again expressed in bar. This expression is based on preliminary results by Parmentier et al. (in preparation) using the method involving passive tracers, and results in values up to 1000 times lower than those obtained with the previous more crude method (see Fig.~\ref{fig:eddy}). In both HD~209458b and HD~189733b we considered a constant $K_{zz}$ value at pressures lower than 10$^{-5}$ bar.

\subsubsection{Dynamical time scales} \label{subsubsec:taudyn}

\begin{figure}
\centering
\includegraphics[angle=0,width=\columnwidth]{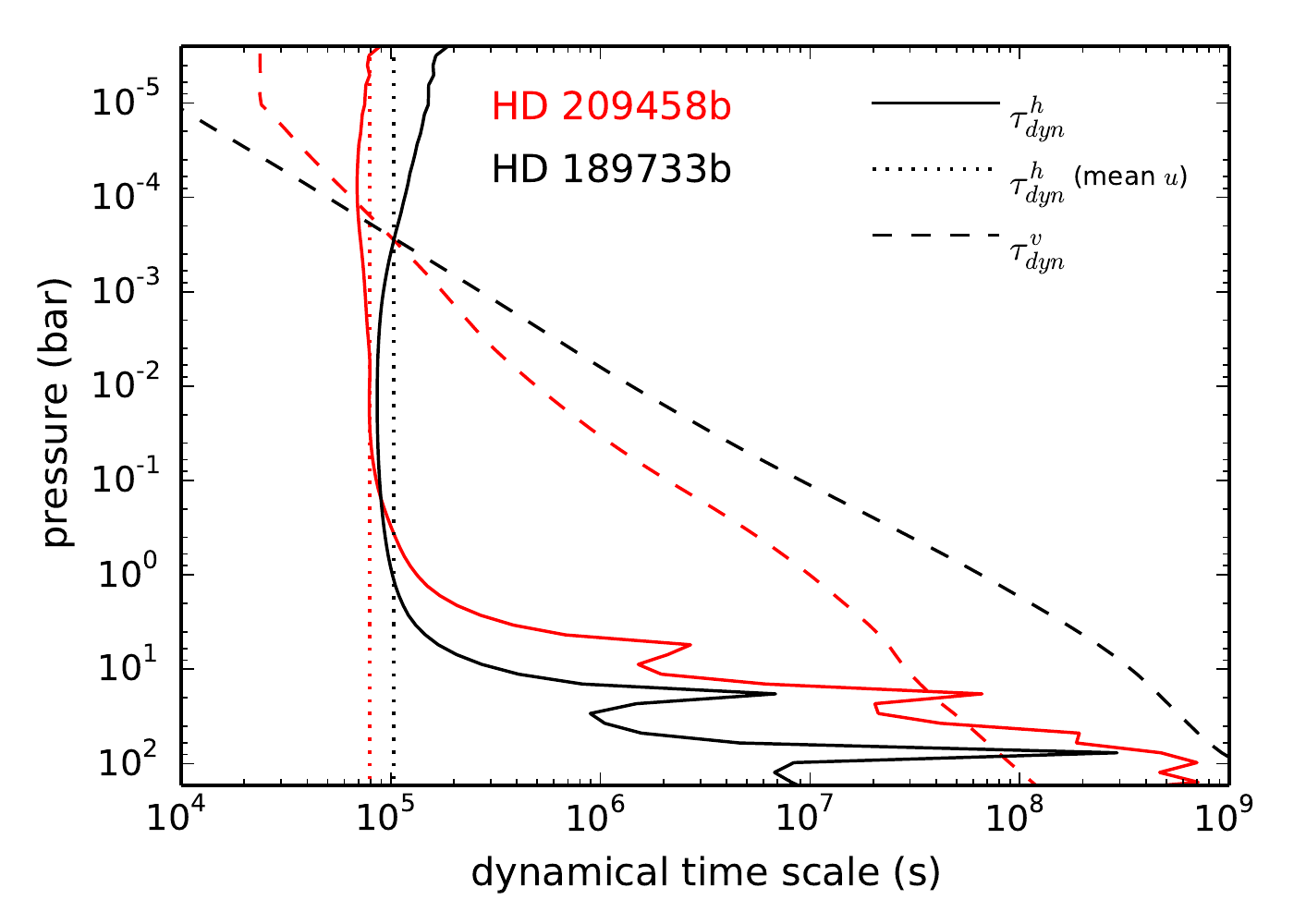}
\caption{Dynamical time scales of horizontal transport ($\tau_{dyn}^{h}$) and vertical mixing ($\tau_{dyn}^{v}$) as a function of pressure in HD~209458b and HD~189733b. $\tau_{dyn}^{h}$ is computed adopting as zonal wind speed either an average over an equatorial band of latitude $\pm$20$^\circ$ (solid lines) or the mean values independent of height given in section~\ref{subsubsec:wind} (dotted lines).} \label{fig:taudyn}
\end{figure}

To assess the relative strengths of horizontal transport and vertical mixing in the atmospheres of HD~209458b and HD~189733b it is useful to argue in terms of dynamical time scales. The dynamical time scale of horizontal transport may be roughly estimated as $\tau_{dyn}^h = \pi R_p / u$, where $R_p$ is the planetary radius and $u$ the zonal wind speed, while that related to vertical mixing can be approximated as $\tau_{dyn}^v = H^2/K_{zz}$, where $H$ is the atmospheric scale height and $K_{zz}$ the eddy diffusion coefficient. If we take a zonal wind speed uniform with altitude and equal to the mean value given in section~\ref{subsubsec:wind}, we find that horizontal transport occurs faster than vertical mixing over most of the vertical structure of the atmospheres of HD~209458b and HD~189733b (see dotted and dashed lines in Fig.~\ref{fig:taudyn}). Only in the upper layers, at pressures below 10$^{-3}$-10$^{-4}$ bar, the high eddy diffusion coefficient makes vertical mixing faster than horizontal transport.

In the deep atmosphere, however, the equatorial superrotating jet vanishes and zonal winds become slower (see Figs.~\ref{fig:uzonal-hd209458b} and \ref{fig:uzonal-hd189733b}), although horizontal transport still remains faster than or at least similar to vertical mixing (see solid and dashed lines in Fig.~\ref{fig:taudyn}). In these deep layers, below the 1-10 bar pressure level, our assumption of a zonal wind speed uniform with altitude and with values as high as a few km s$^{-1}$ is not valid. This is clearly a limitation of the pseudo two-dimensional model, although the implications for the resulting two-dimensional distribution of atmospheric constituents are not strong because in these deep layers the temperature remains rather uniform with longitude, and molecular abundances are largely controlled by thermochemical equilibrium, which makes them quite insensitive to the strength of horizontal transport. This has been verified by running models for HD~209458b and HD~189733b with zonal wind speeds down to 1000 times slower than the nominal mean values given in section~\ref{subsubsec:wind}.\\

In the way the pseudo two-dimensional chemical model is conceived, it clearly deals with the equatorial region of hot Jupiter atmospheres. First, the formalism adopted, in which a vertical atmosphere column rotates around the equator at a constant angular velocity, is adequate for the equatorial region ($\pm$20$^\circ$ in latitude), where a strong eastward jet is found to dominate the circulation according to GCM simulations (see Figs.~\ref{fig:uzonal-hd209458b} and \ref{fig:uzonal-hd189733b}). Second, the temperature structure adopted (see Figs.~\ref{fig:tk-hd209458b} and \ref{fig:tk-hd189733b}) corresponds to the average over an equatorial band of width $\pm$20$^\circ$ in latitude. Third, the rotation period of the atmosphere column is calculated from the wind speed retrieved from the GCM (which is also an average over an equatorial band $\pm$20$^\circ$ in latitude) and the equatorial circumference. And fourth, the longitude-dependent zenith angle adopted to compute the penetration of stellar UV photons corresponds to the equatorial latitude. The adopted formalism is therefore adequate for the equatorial region as long as circulation is dominated by an eastward jet. With these limitations in mind, we now present and discuss the chemical composition distribution resulting from the pseudo two-dimensional chemical model for the atmospheres of HD~209458b and HD~189733b.

\section{Distribution of atmospheric constituents}

\subsection{Overview}

\begin{figure}
\centering
\includegraphics[angle=0,width=\columnwidth]{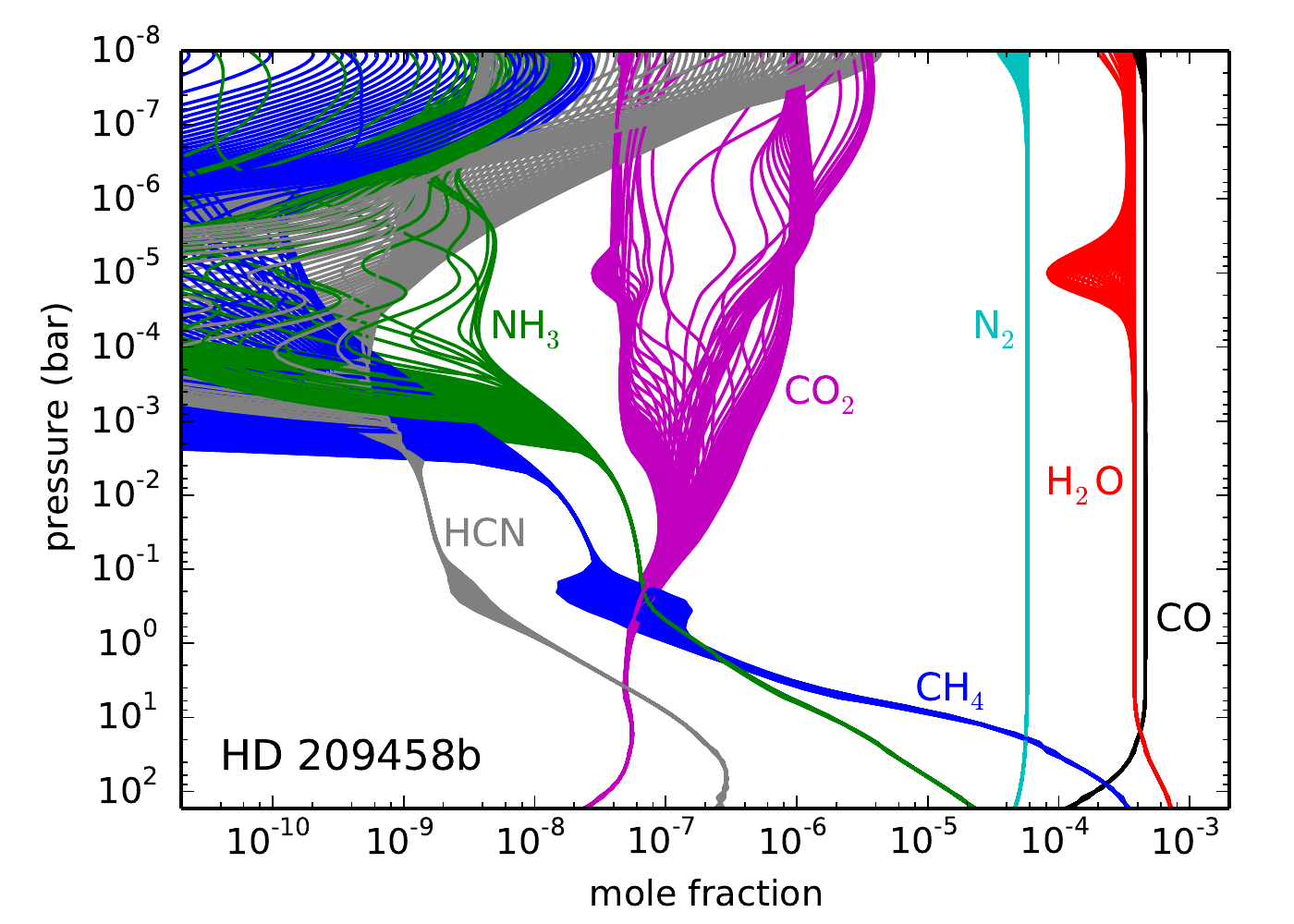}
\caption{Vertical cuts of the abundance distributions of some of the most abundant molecules at longitudes spanning the 0-360$^\circ$ range, as calculated with the pseudo two-dimensional chemical model for HD~209458b's atmosphere.} \label{fig:spaghetti-hd209458b}
\end{figure}

\begin{figure}
\centering
\includegraphics[angle=0,width=\columnwidth]{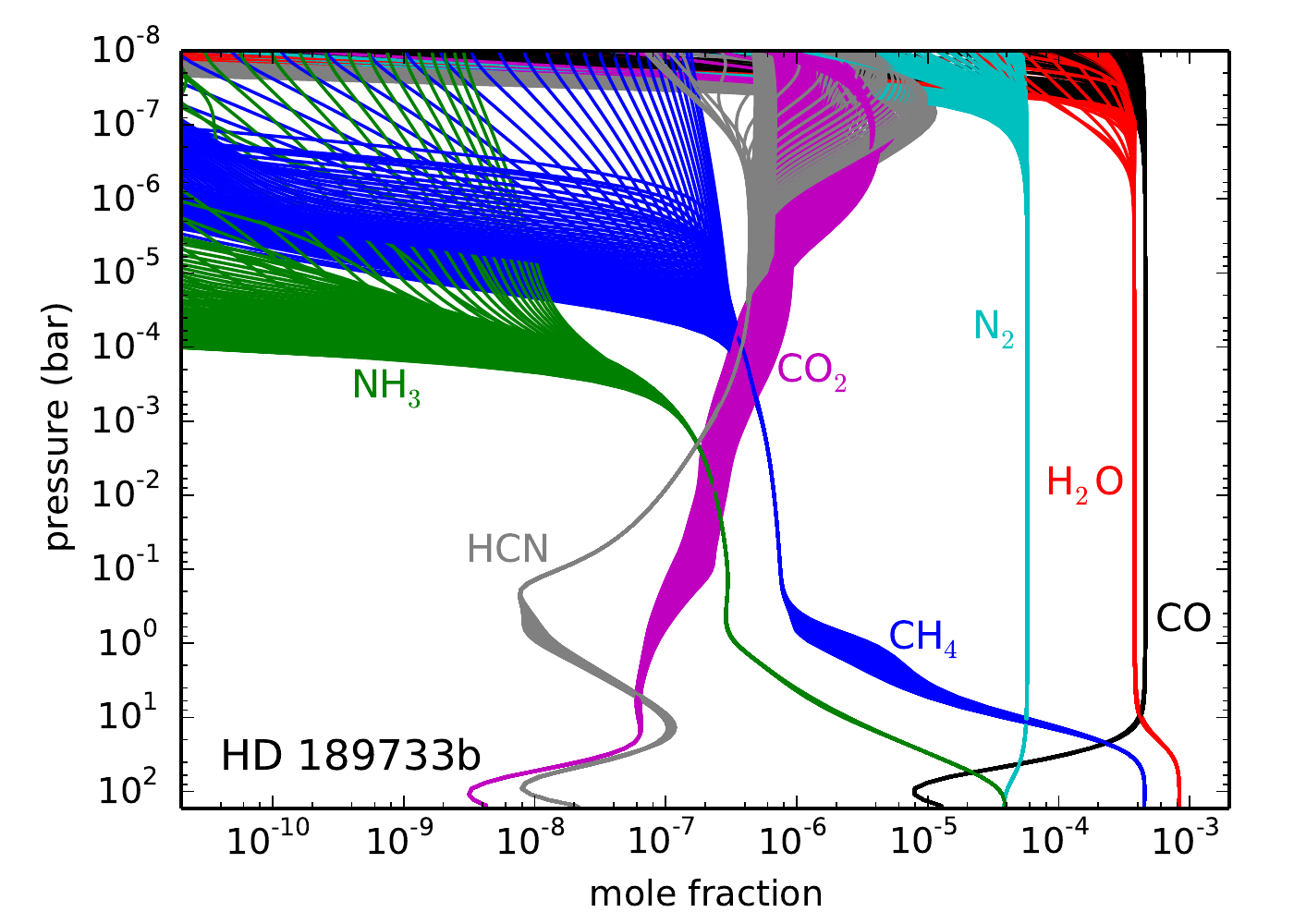}
\caption{Vertical cuts of the abundance distributions of some of the most abundant molecules at longitudes spanning the 0-360$^\circ$ range, as calculated with the pseudo two-dimensional chemical model for HD~189733b's atmosphere.} \label{fig:spaghetti-hd189733b}
\end{figure}

A first glance at the calculated distribution of the chemical composition with altitude and longitude in the atmospheres of HD~209458b and HD~189733b can be obtained by examining the ranges over which the vertical abundance profiles vary with longitude. This information is shown in Figs.~\ref{fig:spaghetti-hd209458b} and \ref{fig:spaghetti-hd189733b} for some of the most abundant species, after H$_2$ and He. We can see that some molecules such as CO, H$_2$O, and N$_2$ show little abundance variation with longitude, while some others such as CH$_4$, CO$_2$, NH$_3$, and HCN experience important changes in their abundances as longitude varies. Abundance variations are usually restricted to the upper regions of the atmosphere (above the 10$^{-1}$-10$^{-3}$ bar pressure level, depending on the molecule) but not to the lower atmosphere, where molecules maintain rather uniform abundances with longitude. On the one hand, longitudinal gradients in the temperature and incident stellar UV flux drive the abundance variations with longitude, while on the other, the zonal wind tends to homogenize the chemical composition in the longitudinal direction, resulting in the complex abundance distributions shown in Figs.~\ref{fig:spaghetti-hd209458b} and \ref{fig:spaghetti-hd189733b}. These results agree with the predictions of \cite{coo2006} (2006) concerning the CO distribution in HD~209458b's atmosphere. These authors coupled a GCM to a simple chemical kinetics scheme dealing with the interconversion between CO and CH$_4$ and found that CO shows a rather homogeneous distribution with longitude and latitude in spite of the strong variations predicted by chemical equilibrium. We also find a rather homogeneous distribution of CO with longitude, although the same is not true for other molecules that display important longitudinal abundance gradients.

\begin{figure}
\centering
\includegraphics[angle=0,width=\columnwidth]{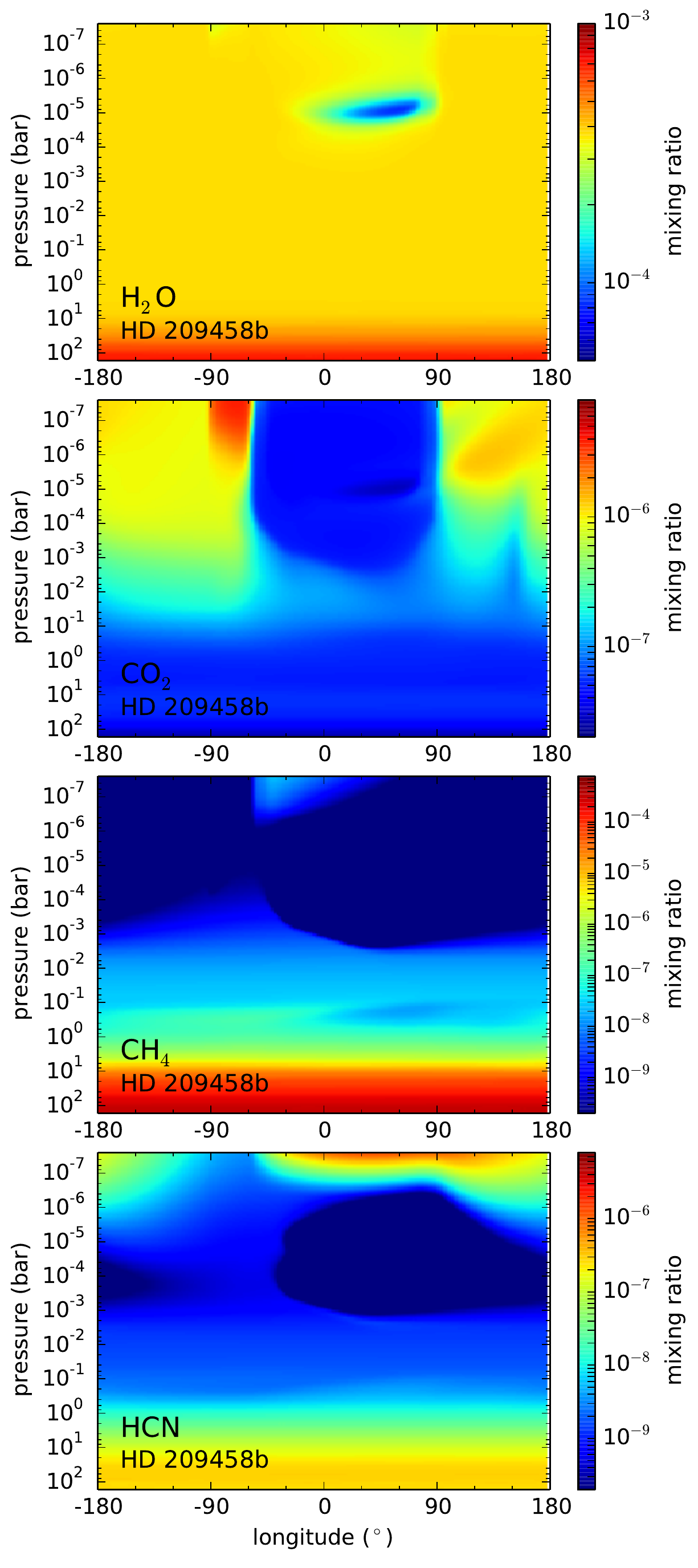}
\caption{Distribution of H$_2$O, CO$_2$, CH$_4$, and HCN as a function of longitude and pressure in the equatorial band of HD~209458b's atmosphere, as calculated with the pseudo two-dimensional chemical model.} \label{fig:abunmap-hd209458b}
\end{figure}

\begin{figure}
\centering
\includegraphics[angle=0,width=\columnwidth]{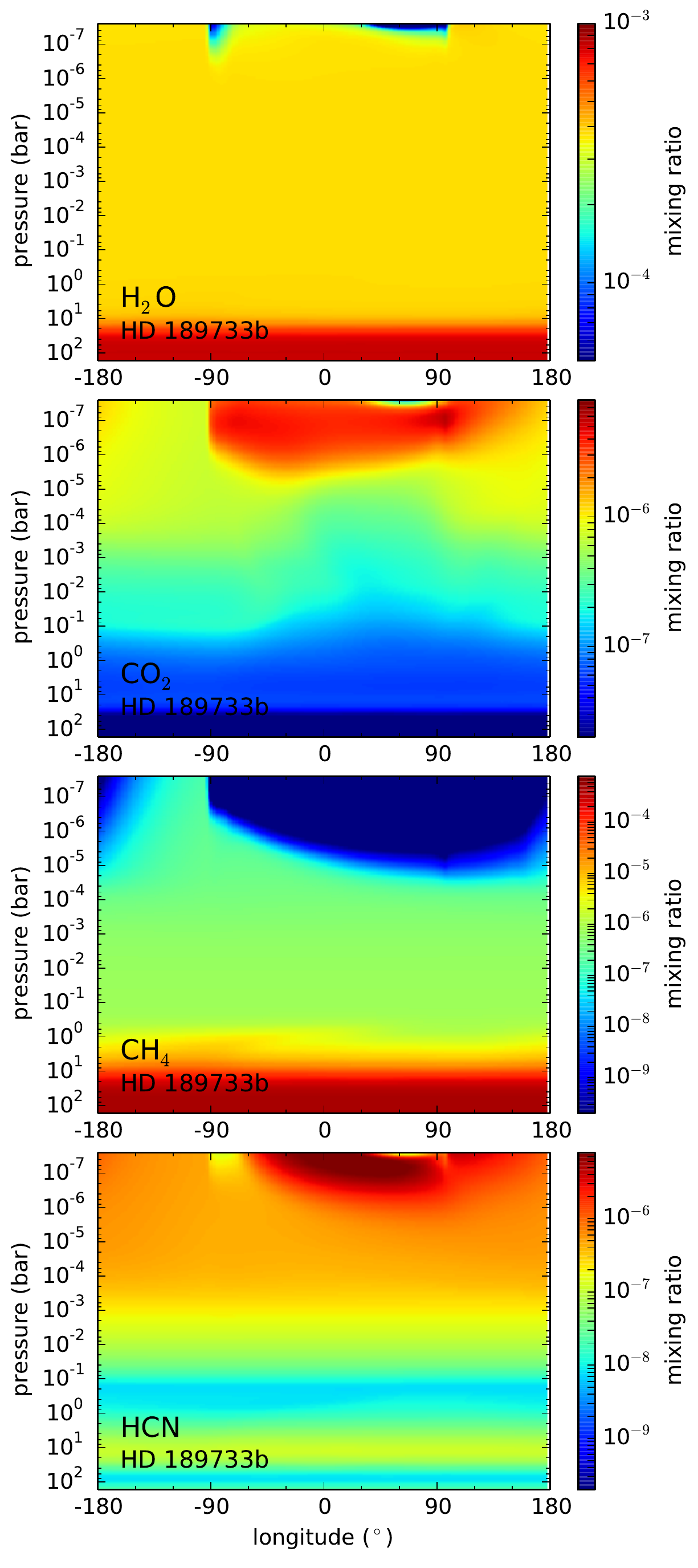}
\caption{Same as Fig.~\ref{fig:abunmap-hd209458b}, but for HD~189733b.} \label{fig:abunmap-hd189733b}
\end{figure}

In Figs.~\ref{fig:abunmap-hd209458b} and \ref{fig:abunmap-hd189733b} we show the atmospheric distribution of selected molecules that may influence planetary spectra as a function of longitude and pressure. Water vapor illustrates the case of a molecule with a rather uniform distribution throughout the atmosphere of both planets, except for a slight enhancement at high pressures ($>$10 bar) and a small depletion, which in HD~209458b occurs at about 10$^{-5}$ bar eastward of the substellar point and is induced by the warm stratosphere, and in HD~189733b takes place in the upper dayside layers (above the 10$^{-7}$ bar pressure level) through photochemical destruction. Carbon monoxide also has a quite uniform distribution and is not shown in Figs.~\ref{fig:abunmap-hd209458b} and \ref{fig:abunmap-hd189733b}. Carbon dioxide is perhaps the most abundant molecule showing important longitudinal abundance variations, with a marked day-to-night contrast. In HD~209458b this molecule is enhanced in the cooler nightside, where it is thermodynamically favored. In HD~189733b the nightside enhancement is only barely apparent in the 10$^{-5}$-10$^{-1}$ bar pressure range, while in upper layers the situation is reversed and CO$_2$ becomes depleted in the nightside regions because of a complex interplay between chemistry and dynamics. In the atmospheres of both planets, CO$_2$ maintains a mixing ratio between a few 10$^{-8}$ and a few 10$^{-5}$. The hydrides CH$_4$ and NH$_3$ show important abundance variations in the vertical direction, their abundance decrease when moving toward upper low-pressure layers, and also some longitudinal variability, which is only important at low abundance levels, however. In HD~209458b, methane is largely suppressed above the 1 mbar pressure level because of the stratosphere. In HD~189733b it is present at a more important level, except in the very upper layers where its depletion in the warmer dayside regions is propagated by the jet to the east, contaminating the nightside regions to a large extent. Hydrogen cyanide also shows important abundance variations with both longitude and altitude. This molecule is greatly enhanced by the action of photochemistry, and thus becomes quite abundant in the upper dayside regions of HD~189733b and to a lower extent in the upper dayside layers of HD~209458b, where photochemistry is largely supressed by the presence of the stratosphere (\cite{mos2011} 2011; \cite{ven2012} 2012). The distribution of HCN in the upper atmosphere shows that the eastward jet results in a contamination of nightside regions with HCN formed in the dayside.

As long as there is an important departure from chemical equilibrium in the atmospheric composition of both planets, the assumption of local chemical equilibrium in the GCM simulations may be an issue and one potential source of inconsistency between the GCM and the chemical model. Much of the thermal budget of these atmospheres, however, is controlled by water vapor, whose abundance is rather uniform and close to chemical equilibrium. This fact may justify to some extent the assumption of local chemical equilibrium in GCMs. We note however that other atmospheric constituents such as CO and CO$_2$ can also play an important role in the thermal balance of hot-Jupiter atmospheres, especially for elemental compositions far from solar, in which case the hypothesis of chemical equilibrium usually adopted in GCMs may not be adequate. Obviously, a more accurate and self-consistent approach would be to couple a robust chemical kinetics network to a GCM, although this is a very challenging computational task.

\subsection{Comparison with limiting cases}

\begin{figure*}
\centering
\includegraphics[angle=0,width=\textwidth]{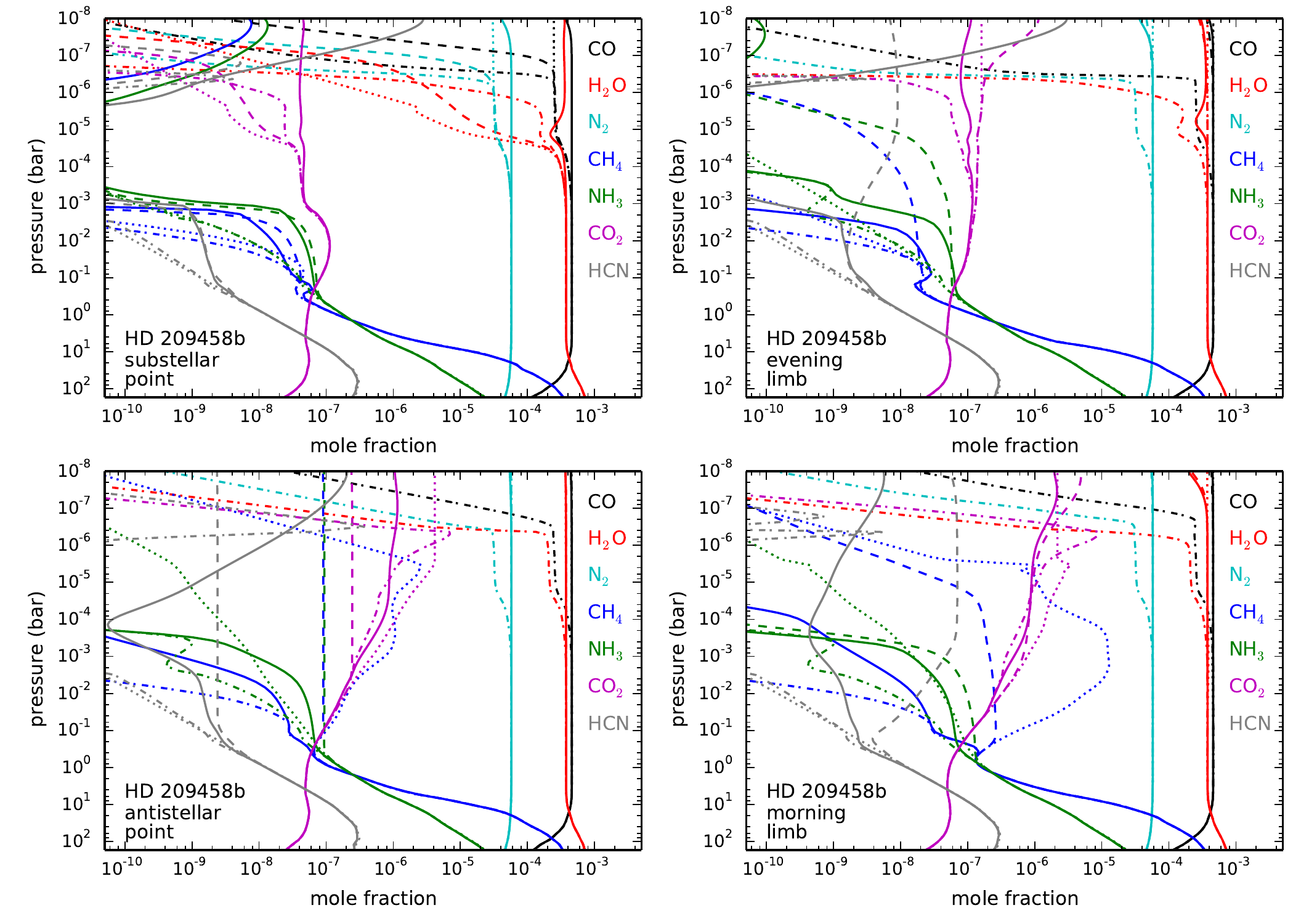}
\caption{Vertical distributions of the most abundant atmospheric constituents, after H$_2$ and He, at four longitudes: substellar point ($0^\circ$), evening limb ($+90^\circ$), antistellar point ($\pm180^\circ$), and morning limb ($-90^\circ$) in the atmosphere of HD~209458b. We show the mole fractions calculated by the pseudo two-dimensional chemical model (solid lines), by a model that neglects vertical mixing (horizontal transport case; dashed-dotted lines), by a one-dimensional vertical model that neglects horizontal transport (vertical mixing case; dashed lines), and by local thermochemical equilibrium (dotted lines). Photochemistry is taken into account in all cases but the last.} \label{fig:abun-hd209458b}
\end{figure*}

\begin{figure*}
\centering
\includegraphics[angle=0,width=\textwidth]{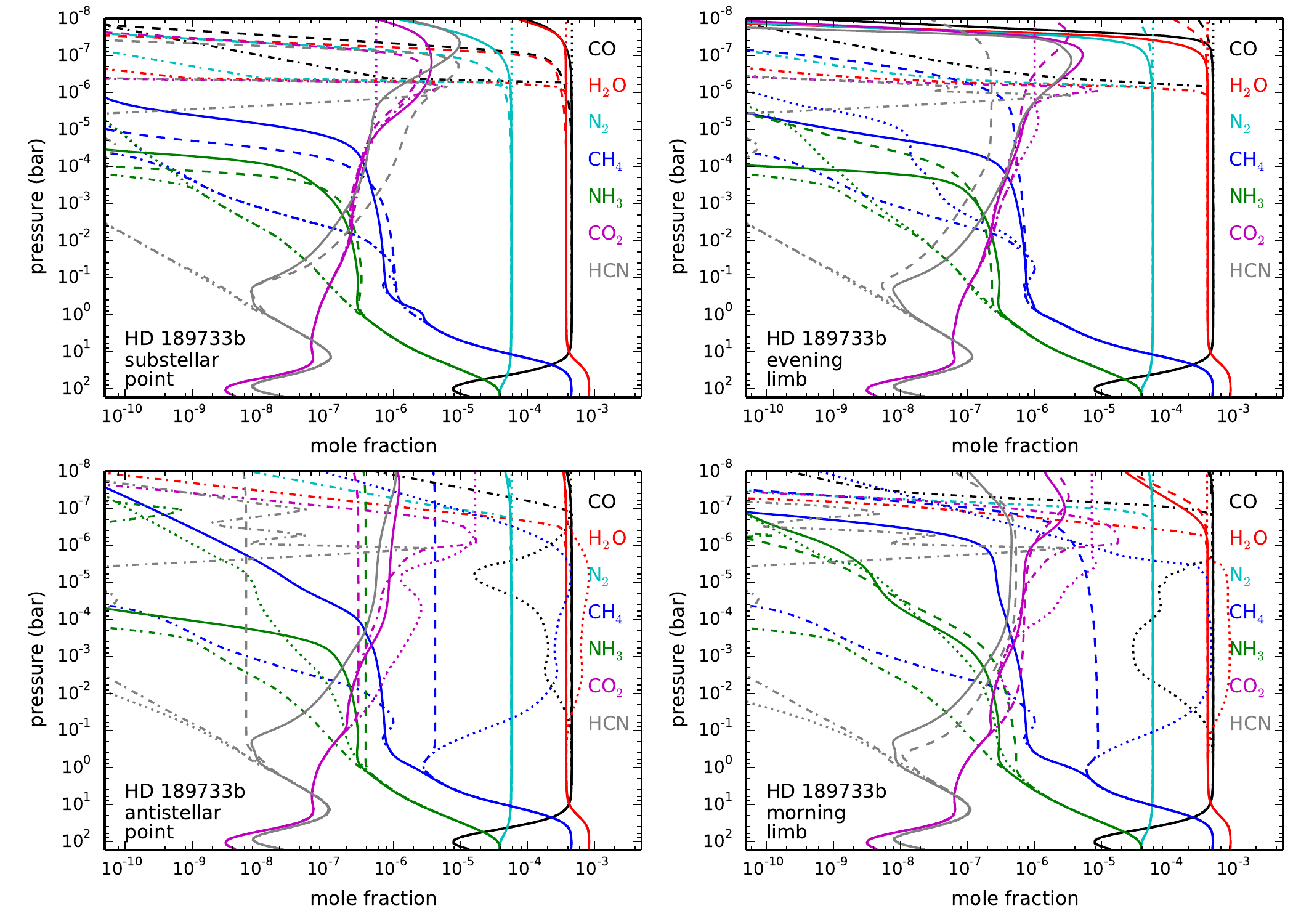}
\caption{Same as Fig.~\ref{fig:abun-hd209458b}, but for HD~189733b.} \label{fig:abun-hd189733b}
\end{figure*}

To obtain insight into the predicted distribution of molecules in the atmospheres of HD~209458b and HD~189733b, a useful and pedagogical exercise is to compare the abundance distributions calculated by the pseudo two-dimensional chemical model with those predicted in various limiting cases. A first one in which vertical mixing is neglected and therefore the only disequilibrium processes are horizontal advection and photochemistry (horizontal transport case), a second one consisting of a one-dimensional vertical model including vertical mixing and photochemistry, which neglects horizontal transport (vertical mixing case), and a third one which is given by local thermochemical equilibrium. Figs.~\ref{fig:abun-hd209458b} and \ref{fig:abun-hd189733b} show the vertical distributions of some of the most abundant species, after H$_2$ and He, in the atmospheres of HD~209458b and HD~189733b, respectively, at four longitudes (substellar and antistellar points, and evening and morning limbs\footnote{We use the terms morning and evening limb to refer to the situation encountered by the traveling wind when crossing each of the two meridians of the planet's terminator. Morning, also called west or leading, and evening, also called east or trailing, limbs are probed by transmission spectra at the ingress and egress, respectively, during primary transit.}), as calculated by the pseudo two-dimensional model and the three aforementioned limiting cases. We may summarize the effects of horizontal transport (modeled as a uniform zonal wind) and vertical mixing (modeled as an eddy diffusion process) by saying that horizontal transport tends to homogenize abundances in the horizontal direction, bringing them close to chemical equilibrium values of the hottest dayside regions, while vertical mixing tends to homogenize abundances in the vertical direction, bringing them close to chemical equilibrium values of hot bottom regions. 

The effect of horizontal transport is perfectly illustrated in the case of methane. In both HD~209458b and HD~189733b, the abundance profile of CH$_4$ given by the horizontal transport case (blue dashed-dotted lines in Figs.~\ref{fig:abun-hd209458b} and \ref{fig:abun-hd189733b}) almost perfectly resembles the chemical equilibrium profile at the substellar point (blue dotted lines in upper left panel of Figs.~\ref{fig:abun-hd209458b} and \ref{fig:abun-hd189733b}), and remains almost invariant with longitude in spite of the important abundance enhancement predicted by chemical equilibrium in the cooler nightside and morning limb regions. The existence of a strong stratosphere in HD~209458b introduces important differences with respect to HD~189733b. The hot temperatures in the upper dayside layers of HD~209458b result in short chemical time scales and therefore allows chemical kinetics to mitigate to some extent the horizontal quenching induced by the zonal wind. This is clearly seen for CO$_2$ (magenta dashed-dotted lines in Figs.~\ref{fig:abun-hd209458b} and \ref{fig:abun-hd189733b}), whose abundance varies longitudinally within 2-3 orders of magnitude in HD~209458b, while in HD~189733b it shows an almost uniform distribution with longitude. The abundance distributions obtained in the horizontal transport case are qualitatively similar to those presented by \cite{agu2012} (2012), although there are some quantitative differences due to the lack of photochemistry and a temperature inversion for HD~209458b in that previous study. Photochemistry plays in fact an important role in the horizontal transport case (dashed-dotted lines in Figs.~\ref{fig:abun-hd209458b} and \ref{fig:abun-hd189733b}), as it causes molecular abundances to vary with longitude in the upper layers due to the switch on/off of photochemistry in the day and night sides, as the wind surrounds the planet. Note also that the lack of vertical mixing in this case causes the photochemically active region to shift down to the level where, in the absence of vertical transport, chemical kinetics is able to counterbalance photodissociations, that is, to synthesize during the night the molecules that have been photodissociated during the day. Another interesting consequence of photochemistry in the horizontal transport case is that molecules such as HCN (gray dashed-dotted lines in Figs.~\ref{fig:abun-hd209458b} and \ref{fig:abun-hd189733b}), which are formed by photochemistry in the upper dayside regions, remain present in the upper nightside regions as a consequence of the continuous horizontal transport, and can in fact increase their abundances through the molecular synthesis ocurring during the night.

In the extreme case where vertical transport completely dominates over any kind of horizontal transport, the homogenization is produced in the vertical, and not longitudinal, direction. The value at which a given molecular abundance is quenched vertically corresponds to the chemical equilibrium abundance at the altitude where the rates of chemical reactions and vertical transport become similar, the so-called quench region. This quench region may be located at a different altitude for each species, although in hot Jupiters such as HD~209458b and HD~189733b it is usually located in the 10-10$^{-2}$ bar pressure range (\cite{mos2011} 2011; \cite{ven2012} 2012; also this study). Assuming the strength of vertical mixing does not vary with longitude (as done in this study), the vertical mixing case would yield uniform abundances with longitude if temperatures do not vary much with longitude in the range of altitudes where abundances are usually quenched vertically. In this case, the quench region for a given species would be the same at all longitudes, and so would the vertically quenched abundance. According to the GCM simulations of HD~209458b and HD~189733b, the temperature varies significantly with longitude above the 1 bar pressure level, and thus the exact values at which the abundances of the different species are quenched vertically vary with longitude. The temperature constrast between day and nightside regions is therefore one of the main causes of abundance variations with longitude, as illustrated by CH$_4$ in both planets (blue dashed lines in Figs.~\ref{fig:abun-hd209458b} and \ref{fig:abun-hd189733b}). Another factor that drives longitudinal abundance gradients in the vertical mixing case is photochemistry, which switches on and off in the day and nightsides, respectively. Without horizontal transport that connects the day and nightsides, abundances become rather flat in the vertical direction in the nightside, where photochemistry is suppressed, and display more complicated vertical abundance profiles in the dayside, where photochemistry causes molecules such as NH$_3$ to be depleted while some others such as HCN are enhanced. Note that because we used eddy diffusion coefficients significantly below those adopted in previous studies (e.g. \cite{mos2011} 2011; \cite{ven2012} 2012), the vertical quench of abundances in the dayside is not as apparent because it is strongly counterbalanced by photochemistry.

In the pseudo two-dimensional model, in which both horizontal transport and vertical mixing are simultaneously taken into account, the distribution of atmospheric constituents (solid lines in Figs.~\ref{fig:abun-hd209458b} and \ref{fig:abun-hd189733b}) results from the combined effect of various processes that tend to drive the chemical composition to a variety of distributions. On the one hand, chemical kinetics proceeds to drive the composition close to local chemical equilibrium. On the other hand, horizontal transport tends to homogenize abundances longitudinally, while vertical mixing does the same in the vertical direction. Finally, stellar UV photons tend to photodissociate molecules in the upper dayside layers, and new molecules are formed through chemical reactions involving the radicals produced in the photodissociations. Among these processes, horizontal transport and vertical mixing compete in homogenizing the chemical composition in the longitudinal and vertical directions, respectively. In the atmospheres of HD~209458b and HD~189733b horizontal transport occurs faster than vertical mixing below the $\sim$1 mbar pressure level (see section~\ref{subsubsec:taudyn}), and therefore molecular abundances are strongly homogenized in the longitudinal direction in this region. In upper layers the competition of mixing and photochemical processes results in a more complex distribution of atmospheric constituents.

Molecular abundances show a wide variety of behaviors when both horizontal transport and vertical mixing are considered simultaneously. The abundances of molecules such as CH$_4$, NH$_3$, and HCN tend to follow those given by the vertical mixing case at the substellar region, but at other longitudes the situation is quite different depending on the molecule (blue, green, and gray solid lines in Figs.~\ref{fig:abun-hd209458b} and \ref{fig:abun-hd189733b}). At the antistellar point, for example, the abundance profiles of CH$_4$ and NH$_3$ are closer to those predicted by the pure horizontal transport case than by the vertical mixing one, but HCN does follow a behavior completely different from each of these two limiting cases. The abundances of CO, H$_2$O, and N$_2$ show little variation with longitude or altitude and are therefore not affected by whether horizontal transport or vertical mixing dominates. Nevertheless, the coupling of horizontal transport and vertical mixing results in some curious behaviors, such as that of water vapor at the substellar point of HD~209458b (red lines in Fig.~\ref{fig:abun-hd209458b}). The two limiting cases of pure horizontal transport and pure vertical mixing predict a decline in its abundance in the upper layers because of photodissociation and because of a low chemical equilibrium abundance at these low pressures. However, horizontal and vertical dynamics working simultaneously bring water from more humid regions so that there is no decline in its abundance up to the top of the atmosphere (at 10$^{-8}$ bar in our model). In summary, taking into account both horizontal transport and vertical mixing produces complex abundance distributions that in many cases cannot be predicted a priori.

\begin{figure}
\centering
\includegraphics[angle=0,width=\columnwidth]{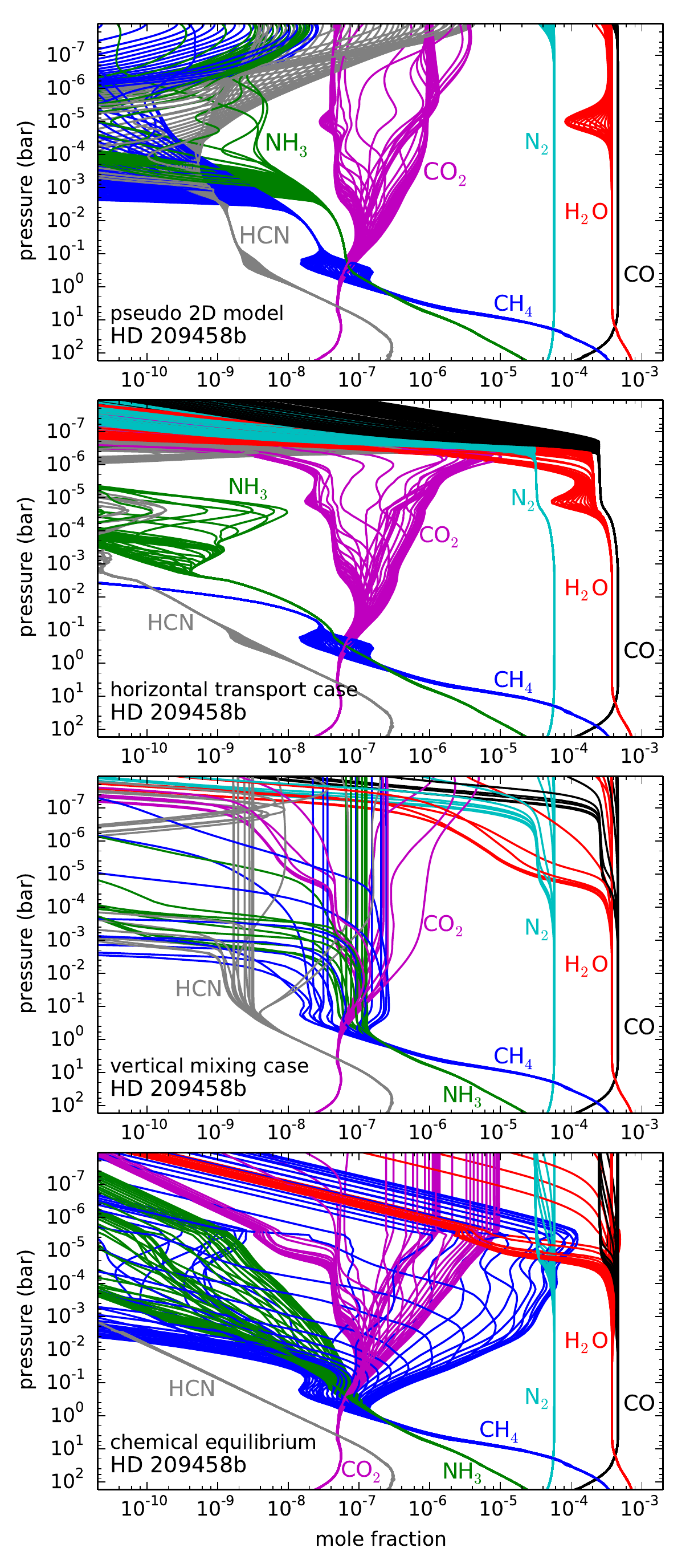}
\caption{Vertical cuts of the abundance distributions of some of the most abundant molecules in HD~209458b's atmosphere at longitudes spanning the 0-360$^\circ$ range, as calculated (from top to bottom) with the pseudo two-dimensional model, in the horizontal transport and vertical mixing cases, and under local chemical equilibrium.} \label{fig:spaghetti-4cases-hd209458b}
\end{figure}

\begin{figure}
\centering
\includegraphics[angle=0,width=\columnwidth]{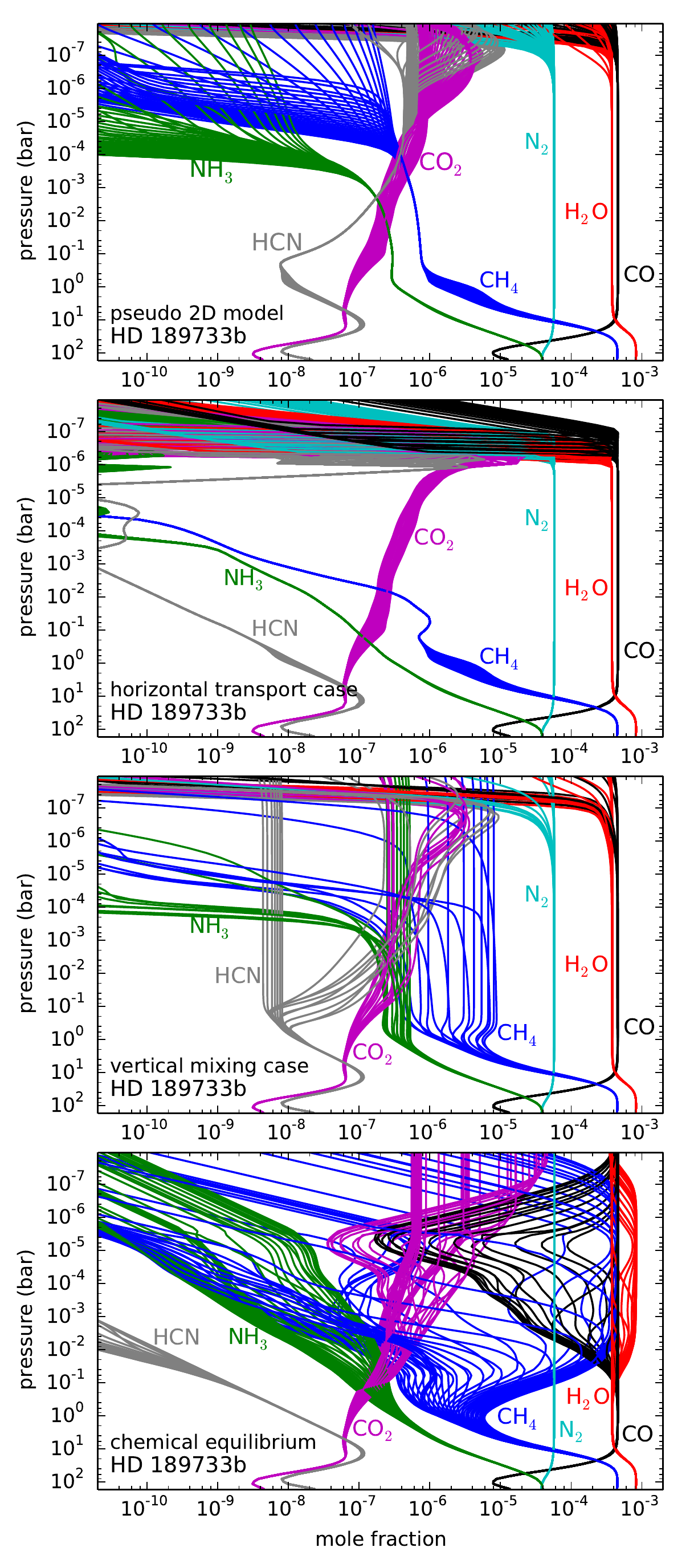}
\caption{Same as Fig.~\ref{fig:spaghetti-4cases-hd209458b}, but for HD~189733b.} \label{fig:spaghetti-4cases-hd189733b}
\end{figure}

We may have a different view of the situation by looking at the ranges over which the vertical abundance profiles vary with longitude in the various limiting cases (see Figs.~\ref{fig:spaghetti-4cases-hd209458b} and \ref{fig:spaghetti-4cases-hd189733b}). Our attention first focuses on the fact that local chemical equilibrium predicts strong variations of the chemical composition with longitude in the atmospheres of both HD~209458b and HD~189733b. This is especially true for CH$_4$ and CO in the latter planet, where methane becomes more abundant than carbon monoxide in the cooler nightside regions. Disequilibrium processes, however, in particular horizontal transport and vertical mixing, reduce to a large extent the longitudinal variability of molecular abundances. As already stated, although perhaps more clearly seen in Figs.~\ref{fig:spaghetti-4cases-hd209458b} and \ref{fig:spaghetti-4cases-hd189733b}, horizontal transport tends to homogenize abundances with longitude. The effect of a purely horizontal transport is perfectly illustrated in HD~189733b's atmosphere, where, except for the photochemically active region in the upper layers, the distribution of molecules is remarkably homogeneous with longitude (see horizontal transport panel in Fig.~\ref{fig:spaghetti-4cases-hd189733b}). In the atmosphere of HD~209458b, on the other hand, a pure horizontal transport allows for some longitudinal variability in the abundances of CO$_2$ and NH$_3$ above the 10$^{-3}$ bar pressure level (see horizontal transport panel in Fig.~\ref{fig:spaghetti-4cases-hd209458b}), mainly because of the activation of chemical kinetics in the dayside stratosphere and its ability to counterbalance the homogenization driven by horizontal transport. In the vertical mixing case (i.e., no horizontal transport), abundances are more uniform in the vertical direction but show important longitudinal variations, with a marked day/night asymmetry characterized by rather flat vertical abundance profiles in the nightside and abundances varying with altitude in the dayside because of the influence of photochemistry (see e.g. CH$_4$, NH$_3$, and HCN in vertical mixing panels of Figs.~\ref{fig:spaghetti-4cases-hd209458b} and \ref{fig:spaghetti-4cases-hd189733b}). When horizontal transport and vertical mixing are considered simultaneously (top panels of Figs.~\ref{fig:spaghetti-4cases-hd209458b} and \ref{fig:spaghetti-4cases-hd189733b}), the distribution of molecules in the lower atmosphere of both HD~209458b and HD~189733b, below the 10$^{-3}$ bar pressure level, remains remarkably homogeneous with longitude and close to that given by the pure vertical mixing case at the substellar regions. That is, the chemical composition of the hottest dayside regions propagates to the remaining longitudes, which indicates that the zonal wind transports material faster than vertical mixing processes do. In the upper atmosphere the abundance profiles become more complicated because of the combined effect of the photochemistry that takes place in the dayside and the mixing of material ocurring in both the vertical and horizontal directions.

\subsection{Comparison with previous one-dimensional models}

\begin{figure}
\centering
\includegraphics[angle=0,width=\columnwidth]{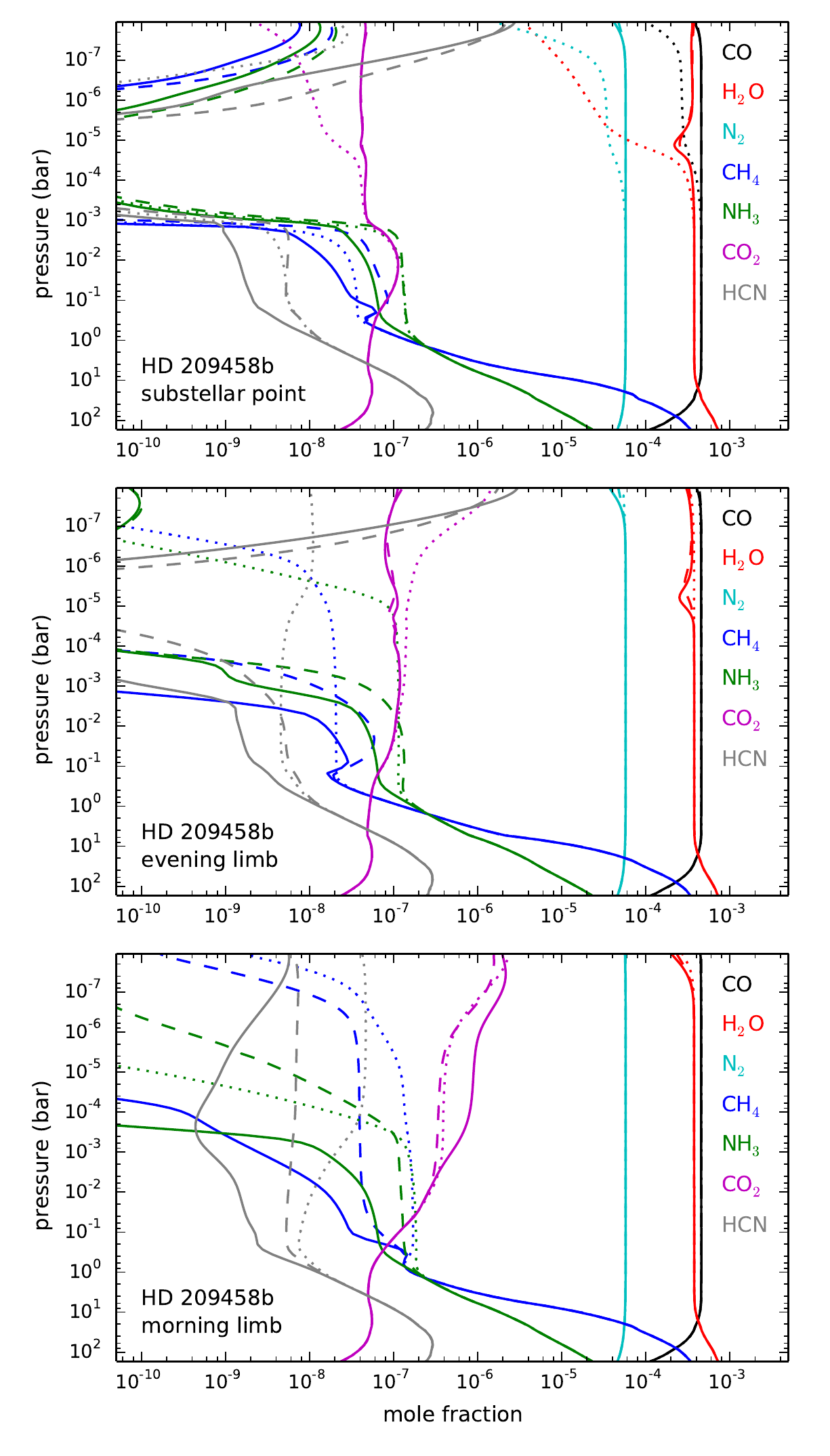}
\caption{Effect of eddy coefficient profile and 1D/2D character of the model on the vertical abundance profiles of some of the most abundant molecules in HD~209458b. We show abundances at the substellar point and at the evening and morning limbs, as given by the pseudo 2D model using the nominal eddy coefficient profile (solid lines), by the pseudo 2D model using the \cite{mos2011} 2011's eddy profile (dashed lines), and by a one-dimensional vertical model using the \cite{mos2011} 2011's eddy profile (dotted lines).} \label{fig:abundances-eddy-hd209458b}
\end{figure}

\begin{figure}
\centering
\includegraphics[angle=0,width=\columnwidth]{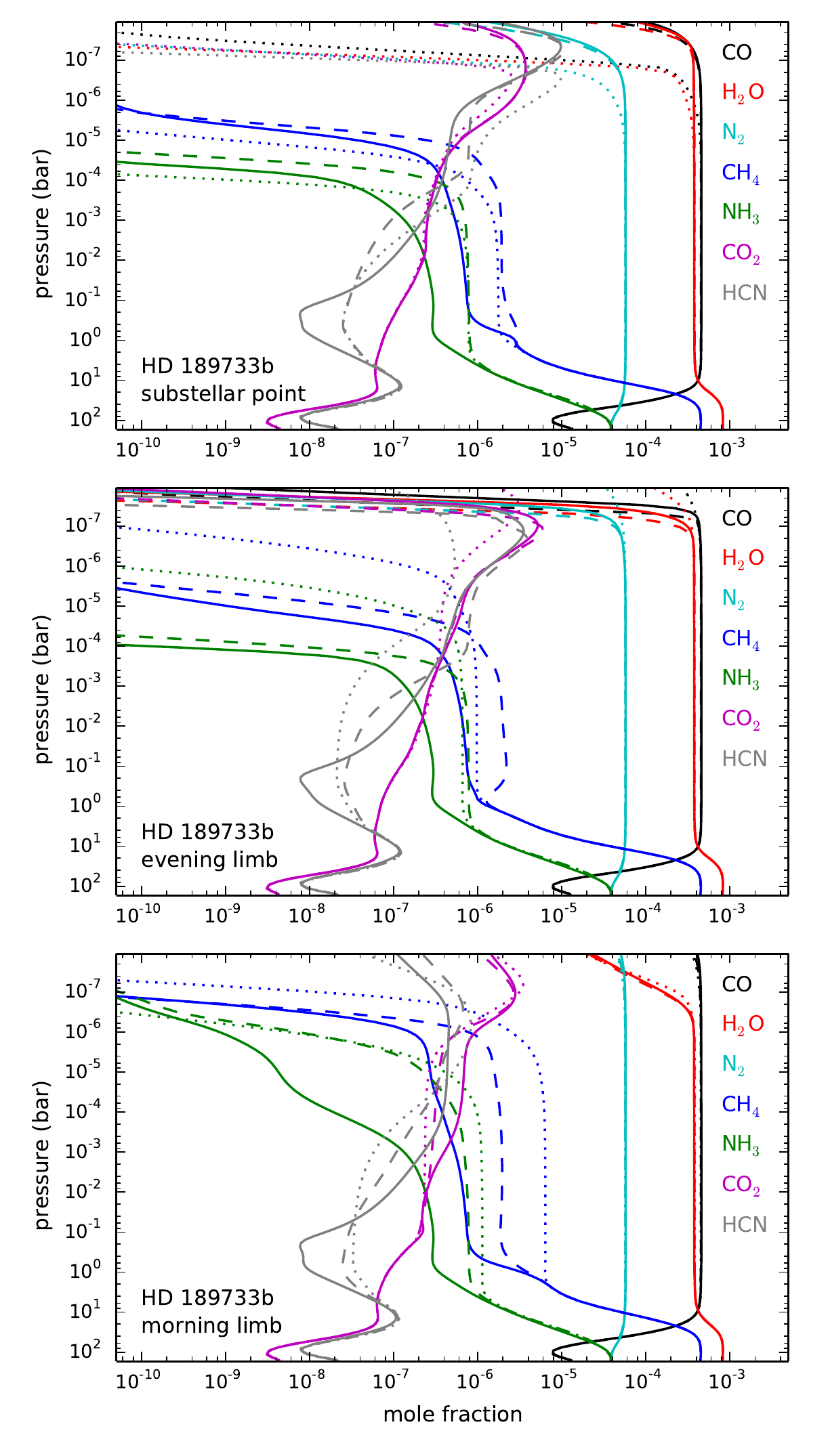}
\caption{Same as Fig.~\ref{fig:abundances-eddy-hd209458b}, but for HD~189733b.} \label{fig:abundances-eddy-hd189733b}
\end{figure}

It is interesting to compare the results obtained with the pseudo two-dimensional model with previous results from one-dimensional vertical models (\cite{mos2011} 2011; \cite{ven2012} 2012). There are two main differences between our model and these previous ones. The first is related to the eddy diffusion coefficients adopted, which are noticeably lower in this study because they are calculated by following the behavior of passive tracers in GCM simulations, while those used previously were also estimated from GCM simulations but as the root mean square of the vertical velocity times the vertical scale height. The second is related to the very nature of the model, which in our case is a pseudo two-dimensional model that simultaneously takes into account horizontal transport and vertical mixing, while in these previous studies horizontal transport is neglected. To isolate the differences caused by each of these factors we compare in Figs.~\ref{fig:abundances-eddy-hd209458b} and \ref{fig:abundances-eddy-hd189733b} the vertical abundance distributions of some of the most abundant molecules at the substellar point and at the two limbs, as calculated with our pseudo two-dimensional model using the nominal vertical profile of the eddy diffusion coefficient (see section~\ref{subsubsec:eddy}), as given by the same pseudo two-dimensional model but using the high eddy diffusion coefficient profiles derived by \cite{mos2011} (2011), which are about 10-100 times higher than ours for HD~209458b and about 10-1000 times higher than ours for HD~189733b, and as computed with a one-dimensional vertical model using the high eddy diffusivity values of \cite{mos2011} (2011).

The main effect of increasing the strength of vertical mixing in the frame of a pseudo two-dimensional model is that the quench region shifts down to lower altitudes. For molecules such as CH$_4$, NH$_3$, and HCN, this implies that their vertically quenched abundances increase (compare solid and dashed lines in Figs.~\ref{fig:abundances-eddy-hd209458b} and \ref{fig:abundances-eddy-hd189733b}). If horizontal transport is completely supressed, that is, moving from a pseudo two-dimensional model to a one-dimensional vertical model (from dashed to dotted lines in Figs.~\ref{fig:abundances-eddy-hd209458b} and \ref{fig:abundances-eddy-hd189733b}), the horizontal homogenization of abundances is completely lost and thus the abundances of species such as CH$_4$ experience more important variations with longitude. Another interesting consequence of suppressing horizontal transport concerns water vapor, carbon monoxide, and molecular nitrogen, whose abundances decrease in the upper layers of the dayside regions (see upper panel in Figs.~\ref{fig:abundances-eddy-hd209458b} and \ref{fig:abundances-eddy-hd189733b}). In HD~209458b the depletion of these molecules is caused by the hot stratosphere, where neutral O and C atoms are favored over molecules, while in HD~189733b it is caused by photodissociation by UV photons. This loss of H$_2$O, CO, and N$_2$ molecules in the upper dayside layers is shifted to higher altitudes when horizontal transport, which brings molecules from other longitudes, is taken into account.

The vertical abundance profiles calculated with the one-dimensional vertical model at the substellar point and at the two limbs (dotted lines in Figs.~\ref{fig:abundances-eddy-hd209458b} and \ref{fig:abundances-eddy-hd189733b}) can be compared with the one-dimensional results of \cite{mos2011} (2011) using averaged thermal profiles for the dayside and terminator regions (see also \cite{ven2012} 2012). There is a good overall agreement between our substellar point results and their dayside results, on the one hand and on the other, between our results at the two limbs and their results at the terminator region, except for CH$_4$, NH$_3$, and HCN, for which we find vertically quenched abundances lower by about one order of magnitude. Although there are some differences in the adopted elemental abundances, stellar UV spectra, and zenith angles, the main source of the discrepancies is attributed to the different temperature profiles adopted. On the one hand there are slight differences between the GCM results of \cite{sho2009} (2009), adopted by \cite{mos2011} (2011) and \cite{ven2012} (2012), and those of \cite{par2013} (2013, in preparation), which are adopted here. On the other, and more importantly, the temperature profiles have a different nature. They are averages over the dayside and terminator regions in their case, while in ours they correspond to specific longitudes. The dayside average temperature profile of \cite{mos2011} (2011) is cooler than our substellar temperature profile by about 100 K around the 1 bar pressure level in both planets, which results in vertically quenched abundances of CH$_4$ and NH$_3$ higher than ours by about one order of magnitude (part of the abundance differences are also due to the different chemical network adopted; see Fig.~7 of \cite{ven2012} 2012). This serves to illustrate how relatively small changes of temperature in the 0.1-10 bar pressure regime --the quench region for most molecules-- may induce important variations in the vertically quenched abundance of certain molecules. This also raises the question of whether it is convenient to use a temperature profile averaged over the dayside in one-dimensional chemical models that aim at obtaining a vertical distribution of molecules representative of the dayside. Although it may be a reasonable choice if one is limited by the one-dimensional character of the model, averaging the temperature over the whole dayside masks the temperatures of the hottest regions, near the substellar point, which are in fact the most important as they control much of the chemical composition at other longitudes if horizontal transport becomes important.

In summary, the main implications of using a pseudo two-dimensional approach and of the downward revision of the eddy values in the atmospheres of HD~209458b and HD~189733b are that, on the the one hand, the longitudinal variability of the chemical composition is greatly reduced compared with the expectations of pure chemical equilibrium or one-dimensional vertical models and, on the other hand, the mixing ratios of CH$_4$, NH$_3$, and HCN are significantly reduced compared with results of previous one-dimensional models (by one order of magnitude or more with respect to the results of \cite{mos2011} 2011), down to levels at which their influence on the planetary spectra are probably minor.

\section{Calculated vs. observed molecular abundances}

We now proceed to a discussion in which we compare the molecular abundances calculated with the pseudo two-dimensional chemical model and those derived from observations. Our main aim here is to evaluate whether or not the calculated composition, which is based on plausible physical and chemical grounds, is compatible with the mixing ratios derived by retrieval methods used to interpret the observations. The molecules H$_2$O, CO, CO$_2$, and CH$_4$ have all been claimed to be detected in the atmospheres of HD~209458b and HD~189733b either in the terminator region of the planet from primary transit observations, in the dayside from secondary eclipse observations, or in both regions using the two methods. Although we are not in a position to cast doubt on any of these detections, given the controversial results often found by different authors in the interpretation of spectra of exoplanets it is advisable to be cautious when using the derived mixing ratios to argue in any direction. Having this in mind, hereafter we use the term detection instead of claim of detection.

Water vapor and carbon monoxide are calculated with nearly their maximum possible abundances in both planets and show a rather homogeneous distribution as a function of both altitude and longitude (see Figs.~\ref{fig:spaghetti-hd209458b} and \ref{fig:spaghetti-hd189733b}). Adopting a solar elemental composition, as done here, the calculated mixing ratios of both H$_2$O and CO are around 5 $\times$ 10$^{-4}$. Water vapor being the species that provides most of the atmospheric opacity at infrared wavelengths, it was the first molecule to be detected in the atmosphere of an extrasolar planet, concretely in the transmission spectrum of HD~189733b (\cite{tin2007} 2007), and H$_2$O mixing ratios derived from observations for both HD~189733b and HD~209458b are usually in the range of the calculated value of 5 $\times$ 10$^{-4}$ (\cite{tin2007} 2007; \cite{gri2008} 2008; \cite{swa2008} 2008, 2009a,b; \cite{mad2009} 2009; \cite{bea2010} 2010; \cite{lee2012} 2012; \cite{lin2013} 2013; \cite{deming2013} 2013). Carbon monoxide, although less evident than water vapor, has also been detected in both planets and the mixing ratios derived are in the range of the values inferred for H$_2$O and expected from the chemical model (\cite{swa2009a} 2009a; \cite{des2009} 2009; \cite{mad2009} 2009; \cite{lee2012} 2012; \cite{lin2013} 2013).

The calculated mixing ratio of carbon dioxide in the two hot Jupiters is in the range 10$^{-7}$ - 10$^{-6}$ depending on the pressure level, with a more important longitudinal variation in the atmosphere of HD~209458b than in that of HD~189733b (see Figs.~\ref{fig:spaghetti-hd209458b} and \ref{fig:spaghetti-hd189733b}). This molecule has been also detected through secondary-eclipse observations in the dayside of HD~189733b, with mixing ratios spanning a wide range from 10$^{-7}$ up to more than 10$^{-3}$ (\cite{swa2009a} 2009a; \cite{mad2009} 2009; \cite{lee2012} 2012; \cite{lin2013} 2013), and in the dayside of HD~209458b, with a mixing ratio in the range 10$^{-6}$ - 10$^{-5}$ (\cite{swa2009b} 2009b). Taking into account the uncertainties associated with the values retrieved from observations, the agreement with the calculated abundance is reasonably good for CO$_2$.

The most important discrepancies between calculated and observed abundances are probably found for methane. This molecule is predicted to be very abundant in the cooler nightside regions of both planets, especially in HD~189733b, according to chemical equilibrium (see lower panels in Figs.~\ref{fig:spaghetti-4cases-hd209458b} and \ref{fig:spaghetti-4cases-hd189733b}), but reaches quite low abundances everywhere in the atmosphere according to the pseudo two-dimensional non-equilibrium model (see upper panels in Figs.~\ref{fig:spaghetti-4cases-hd209458b} and \ref{fig:spaghetti-4cases-hd189733b}). In both hot Jupiters, the calculated mixing ratio of CH$_4$ is in fact significantly lower than the predictions of previous one-dimensional models (\cite{mos2011} 2011; \cite{ven2012} 2012), a finding that strengthens the conflict with observations. We find that the mixing ratio of CH$_4$ above the 1 bar pressure level is below 10$^{-7}$ in HD~209458b and below 10$^{-6}$ in HD~189733b, whatever the side of the planet.

In HD~209458b, secondary-eclipse observations have been interpreted as evidence of methane being present in the dayside with a mixing ratio between 2 $\times$ 10$^{-5}$ and 2 $\times$ 10$^{-4}$ (\cite{swa2009b} 2009b), or within the less constraining range 4 $\times$ 10$^{-8}$ - 3 $\times$ 10$^{-2}$ (\cite{mad2009} 2009). In fact, the abundance of CH$_4$ retrieved in these studies is similar or even higher than that retrieved for H$_2$O, which is clearly not the case according to our predictions. It seems difficult to reconcile the low abundance of CH$_4$ calculated by the pseudo two-dimensional chemical model with the high methane content inferred from observations, which points to some fundamental problem in either of the two sides. As concerns the chemical model, an enhancement of the vertical transport to the levels adopted by \cite{mos2011} (2011) or the supression of horizontal transport would increase the abundance of CH$_4$ only slightly (see Fig.~\ref{fig:abundances-eddy-hd209458b}). Photochemistry, which might potentially enhance the abundance of CH$_4$, is largely supressed by the stratosphere in the dayside atmosphere of HD~209458b. An elemental composition of the planetary atmosphere far from the solar one with, for example, an elemental C/O abundance ratio higher than 1, or some unidentified disequilibrium process, which might be related to, for instance, clouds or hazes, might lead to a high methane content in the warm atmospheric layers of HD~209458b's dayside. Some problems on the observational side cannot be ruled out, taking into account the difficulties associated to the acquisition of photometric fluxes of exoplanets and the possibility of incomplete spectroscopic line lists of some molecules relevant to the interpretation of spectra of exoplanets (see e.g. the recently published line list for hot methane by \cite{har2012} 2012).

In HD~189733b, contradictory results exist on the detection of methane in both the terminator and dayside regions. \cite{swa2008} (2008) reported the detection of CH$_4$ through primary-transit observations, with a derived mixing ratio of about 5 $\times$ 10$^{-5}$, although \cite{sin2009} (2009) did not find evidence of its presence in the transmission spectrum. These contradictory results obtained using NICMOS data could point to non-negligible systematics in the data (e.g., \cite{gib2012} 2012). Controversial results also exist on the detection of CH$_4$ in the dayside emission spectrum of HD~189733b (\cite{swa2009a} 2009a, 2010; \cite{mad2009} 2009; \cite{wal2012} 2012; \cite{lee2012} 2012; \cite{lin2013} 2013; \cite{bir2013} 2013). Until observations can draw more reliable conclusions it is difficult to decide whether or not observations and models are in conflict regarding the abundance of CH$_4$ in HD~189733b.

\section{Variations in the planetary spectra}

\begin{figure}
\centering
\includegraphics[angle=0,width=\columnwidth]{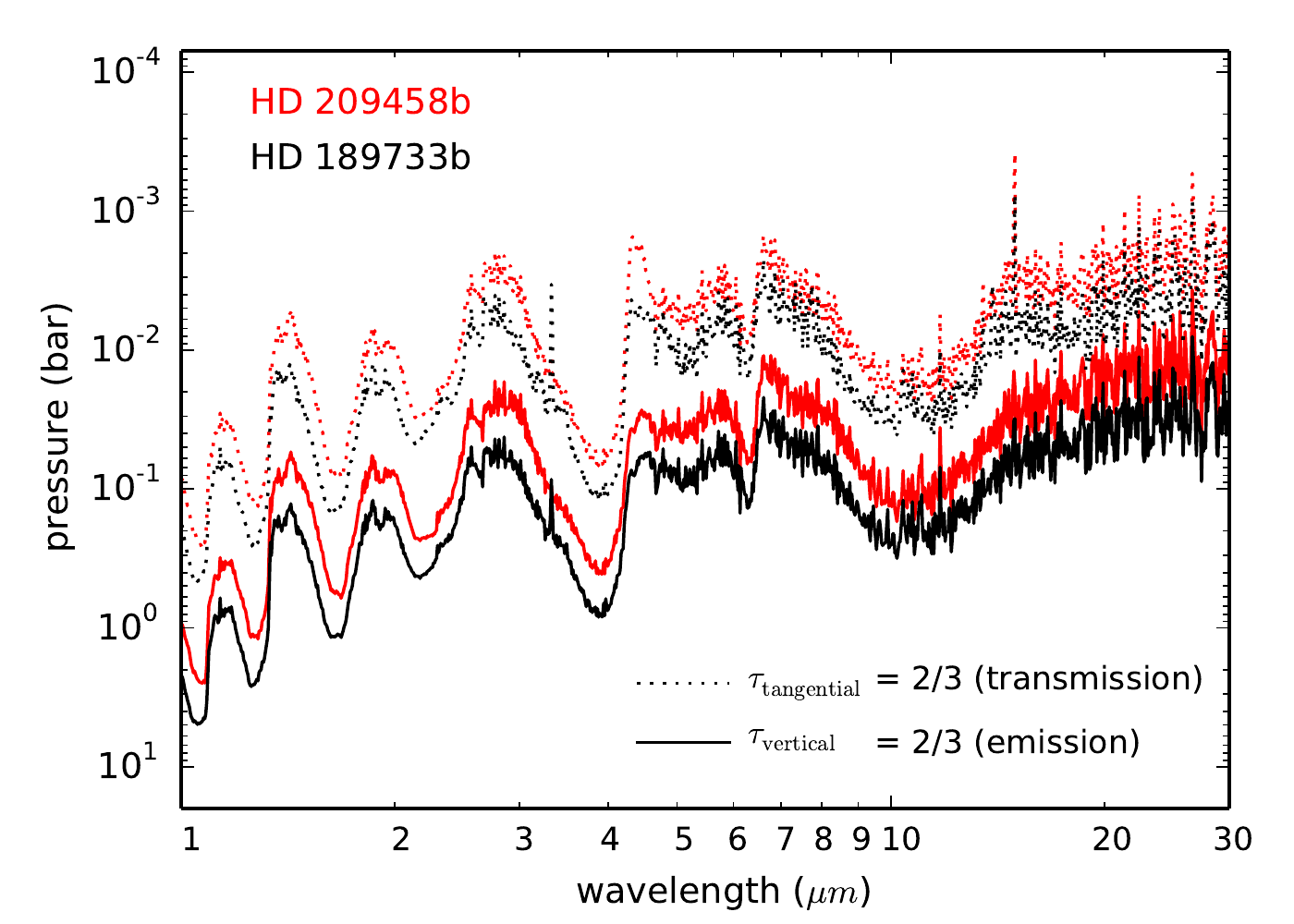}
\caption{Pressure level probed by transmission and emission spectra as a function of wavelength for HD~209458b and HD~189733b. Dashed lines correspond to a model with a mean vertical profile averaged over the terminator and show the pressure level at which the tangential optical depth equals 2/3, which is in fact a transmission spectrum expressed in terms of atmospheric pressure instead of planetary radius. Solid lines correspond to a model with a mean vertical profile averaged over the dayside and indicate the pressure level at which the optical depth in the vertical outward direction equals 2/3, which is an approximate location of the region from where most of the planetary emission arises, the regions below being opaque and those above being translucent.} \label{fig:tau}
\end{figure}

\begin{figure*}
\centering
\includegraphics[angle=0,width=\textwidth]{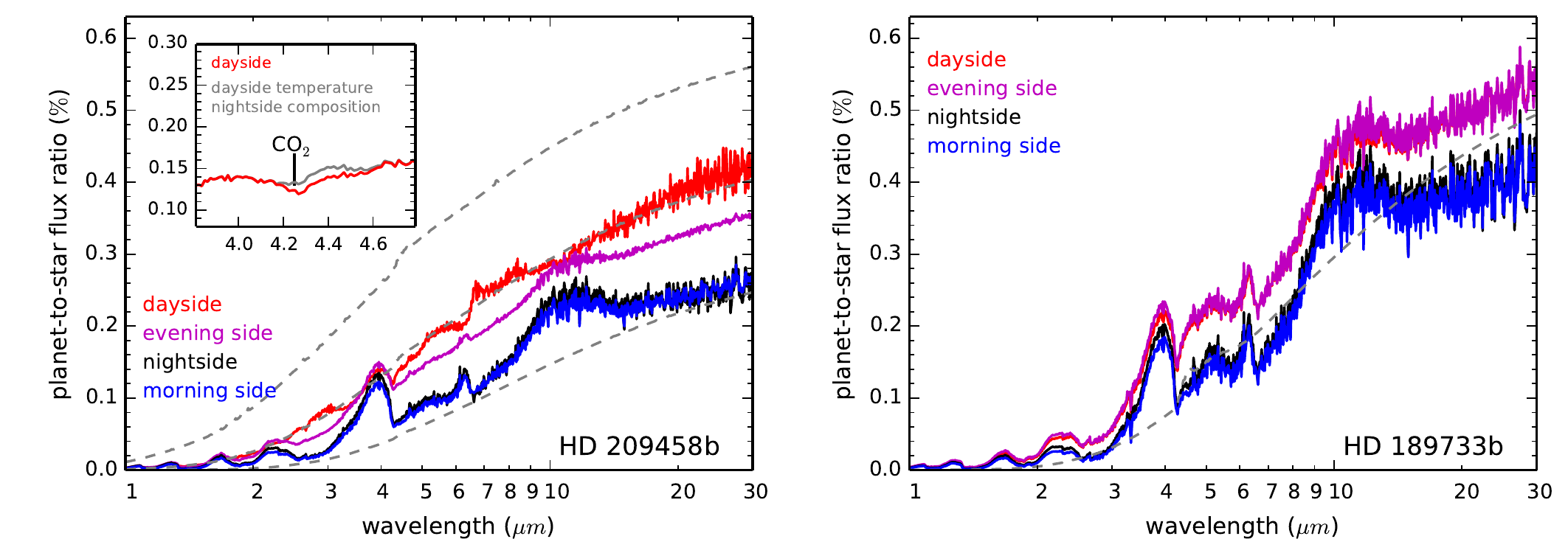}
\caption{Calculated emission spectra for HD~209458b (left panel) and HD~189733b (right panel) at 4 different phases in which the planet faces the observer the day and night sides, and the two sides in between, centered on the evening and morning limbs. Spectra have been smoothed to a resolving power $R$ = 300. The adopted vertical profiles of temperature and mixing ratios are projected area-weighted averages over the corresponding emitting hemisphere, with the distribution of molecules given by our nominal pseudo two-dimensional chemical model. The planetary flux is shown relative to that of the star, for which the Kurucz synthetic spectrum and stellar radius given in section~\ref{sec:model} are adopted. Gray dashed lines correspond to planetary blackbody temperatures of 2000, 1500, and 1000 K (from top to bottom in HD~209458b's panel) and of 1000 K (in HD~189733b's panel). The inset in HD~209458b's panel compares the emission spectrum around 4.3 $\mu$m as calculated for the dayside and as computed using the mean dayside temperature profile and the mean chemical composition of the nightside. The two spectra are nearly identical except for a slight difference at 4.3 $\mu$m due to CO$_2$.} \label{fig:emission}
\end{figure*}

The calculated distribution of molecules in the atmospheres of HD~209458b and HD~189733b may be probed by observations. Instead of comparing synthetic spectra and available observations of these two planets, we are here mainly interested in evaluating whether the longitudinal variability of the chemical composition may be probed by observations. For example, the monitoring of the emission spectrum of the planet at different phases during an orbital period would probe the composition in the different sides of the planet. In addition, the observation of the transmission spectrum at the ingress and egress during primary transit conditions would allow one to probe possible chemical differentiation between the morning and evening limbs of the planet's terminator. Planetary emission and transmission spectra were computed using the line-by-line radiative transfer code described in Appendix~\ref{app:spectra}. Since the code is currently limited because it is one-dimensional in the vertical direction, we adopted mean vertical profiles by averaging the temperature structure in longitude and latitude given by the GCM simulations of \cite{par2013} (2013, in preparation) and the longitudinal distribution of abundances obtained with the pseudo two-dimensional chemical model. In the case of emission spectra, we adopted a weighted average profile of temperature and of mixing ratios over the hemisphere facing the observer (weighted by the projected area on the planetary disk to better represent the situation encountered by an observer), where mixing ratios were assumed to be uniform with latitude. In transmission spectra, vertical profiles are simply obtained by averaging over the whole terminator, or over the morning or evening limb. After adopting an average pressure-temperature profile, the planetary radius (see values in section~\ref{sec:model}) is assigned to the 1 bar pressure level and the altitude of each layer in the atmosphere is computed according to hydrostatic equilibrium. We note that similarly to the case of one-dimensional chemical models, the use of average vertical profiles in calculating planetary spectra is an approximation that masks the longitudinal and latitudinal structure of temperature and chemical composition and may result in non-negligible inaccuracies in the appearance of the spectra, which we plan to investigate in the future.

Is is useful to begin our discussion on planetary spectra with a pedagogical plot that shows the pressure level probed by transmission and emission spectra for HD~209458b and HD~189733b (see Fig.~\ref{fig:tau}). We first note that transmission and emission spectra probe different pressure levels, with transmission spectra being sensitive to upper atmospheric layers than emission spectra. At infrared wavelengths (1-30 $\mu$m), and for the thermal and chemical composition profiles adopted by us for these two hot Jupiters, emission spectra probe pressures between 10 and 10$^{-2}$ bar, while transmission spectra probes the 1-10$^{-3}$ bar pressure regime. A second aspect worth noting is that there are strong variations with wavelength in both types of spectra, which implies that observations at different wavelengths are sensitive to the physical and chemical conditions of different pressure levels. It is always useful to keep these ideas in mind when analyzing the vertical distribution of molecules calculated with a chemical model, because only a very specific region of the atmosphere becomes relevant to planetary spectra.

\subsection{Variation of emission spectra with phase}

A modulation of the planetary emission with the orbital phase has been observed for HD~189733b by monitoring the photometric flux in the 8 $\mu$m band of Spitzer IRAC during a good part of the orbit of the planet (\cite{knu2007} 2007). This has served to evidence the important temperature contrast between the different sides of the planet, noticeably between day and night, and indirectly the presence of strong winds that can redistribute the energy from the day to the night side, due to an observed shift between the hot spot and the substellar point. Various theoretical studies have also been interested in predicting the variation of the planetary flux with the orbital phase in HD~209458b and HD~189733b using the temperature structure calculated with GCM simulations (\cite{for2006} 2006; \cite{sho2008} 2008, 2009; \cite{rau2008} 2008; \cite{bur2010} 2010; \cite{rau2013} 2013). Most previous studies have focused on the link between light curves and variations of temperature between the different planetary sides, and on the comparison between predicted and observed photometric fluxes. Here we are instead interested in discussing the influence of the temperature but also that of the chemical composition (assumed to be given by chemical equilibrium in previous studies) on the variation of the planetary emission with phase.

\begin{figure*}
\centering
\includegraphics[angle=0,width=\textwidth]{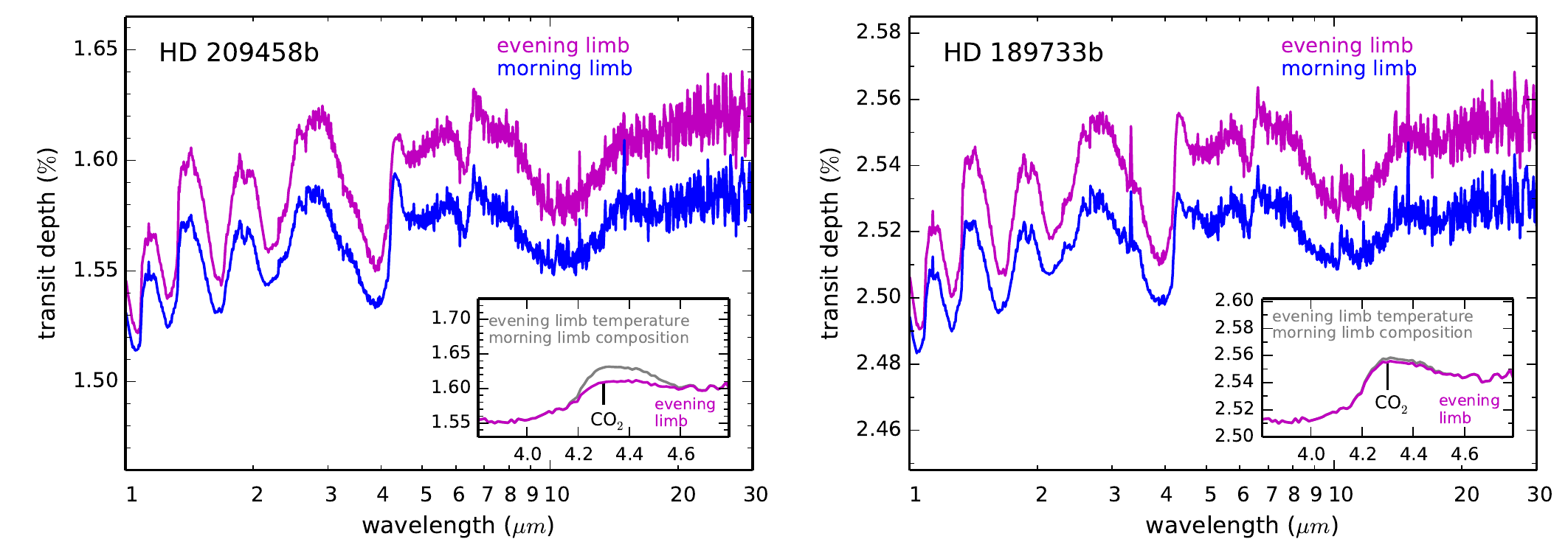}
\caption{Calculated transmission spectra for the evening and morning limbs of HD~209458b (left panel) and HD~189733b (right panel), where the vertical structure is obtained by averaging the temperature over each limb and adopting the abundance profiles at each limb from the nominal pseudo two-dimensional chemical model. Spectra have been smoothed to a resolving power $R$ = 300. The transit depth is simply calculated as $(R_p (\lambda) / R_*)^2$, where $R_p (\lambda)$ is the calculated radius of the planet as a function of wavelength and $R_*$ is the stellar radius (see values in section~\ref{sec:model}). The absolute scale of the transmission spectrum is set by our choice of assigning the value of the planetary radius given in section~\ref{sec:model} to the 1 bar pressure level. Since no attempt has been made to reproduce the absolute scale indicated by primary transit observations, calculated transit depths are somewhat higher than given by observations. The insets in both panels compare the transmission spectrum around 4.3 $\mu$m as calculated for the evening limb and as computed using the mean temperature profile of the evening limb and the chemical composition corresponding to the morning limb. The most important differences between both spectra occur around 4.3 and 15 $\mu$m, due to CO$_2$.} \label{fig:transmission}
\end{figure*}

We show in Fig.~\ref{fig:emission} how the calculated emission spectra of HD~209458b and HD~189733b vary with the phase of the planet. Important variations with phase are apparent in HD~209458b, whose strong dayside stratosphere causes the dayside emission spectrum to be significantly brighter and to have a noticeably different spectral shape than at other phases. In HD~189733b, the modulation of the flux and the variation of the spectral shape with phase are also important although less pronounced. Emission spectra are controlled on the one hand, by the vertical temperature structure, and on the other, by the abundances of the main atmospheric constituents providing opacity. The sensitivity of emission spectra to the thermal structure is illustrated in HD~209458b, whose dayside (facing a temperature inversion to the observer) shows some spectral features that appear in emission and not in absorption, as occurs for the other planetary sides of HD~209458b and for HD~189733b. These differences in the spectra can be used to infer whether there is a stratosphere in the atmosphere of a hot Jupiter from observations of its dayside emission spectrum (\cite{knu2008} 2008, 2009b; \cite{bur2008} 2008; \cite{mac2008} 2008, 2010; \cite{tod2010} 2010, \cite{mad2010} 2010). It is also interesting to note how similar the emission spectra of night and morning sides are in the two planets, as are the day and evening sides in the case of HD~189733b. This is a consequence of the eastward transport of energy by the superrotating jet, which shifts the hottest and coldest regions to the east of the substellar and antistellar points, respectively. In the calculated emission spectra of both HD~209458b and HD~189733b, most of the atmospheric opacity along the 1-30 $\mu$m wavelength range is provided by water vapor, with carbon monoxide contributing at 2.3 and 4.6 $\mu$m, CO$_2$ at 4.3 and 15 $\mu$m, and collision-induced absorption by H$_2$-H$_2$ in certain wavelength ranges below 4 $\mu$m. No other species leaves appreciable signatures in the calculated emission spectra of HD~209458b, although in that of HD~189733b CH$_4$ contributes around 3.3 and 7.7 $\mu$m, NH$_3$ around 10.6 $\mu$m, and HCN at 14 $\mu$m.

An interesting question that arises from the change in the emission spectrum with phase is whether it is entirely caused by the variation of temperature in the different sides of the planet or whether the longitudinal variation of the chemical composition contributes to an important extent. To illustrate this point we compare in the inset of HD~209458b's panel in Fig.~\ref{fig:emission} the dayside emission spectrum with a synthetic spectrum calculated using the mean vertical temperature structure of the dayside and the mean chemical composition of the nightside. The two spectra are nearly identical except for a slight difference around 4.3 $\mu$m, a spectral region where atmospheric opacity is to a large extent dominated by CO$_2$. Similar models in which the temperature structure and the chemical composition are adopted from different planetary sides indicate that variations of HD~209458b's emission spectrum with phase are almost entirely caused by changes of temperature, with the only effect that can be purely adscribed to variations in the chemical composition being restricted to the tiny variation (less than 0.02 \% in the planet-to-star flux ratio) at 4.3 $\mu$m, which is caused by the longitudinal variation of about one order of magnitude in the abundance of CO$_2$ (see Fig.~\ref{fig:spaghetti-hd209458b}). The reasons of the small impact of the chemical composition on the variation of emission spectra with phase are related to the important longitudinal homogenization of the abundances driven by the zonal wind in HD~209458b (see Fig.~\ref{fig:spaghetti-hd209458b}). In fact, most of the atmospheric opacity affecting the emission spectrum comes from H$_2$O, CO, and CO$_2$, in order of decreasing importance, and the two former molecules show remarkably uniform abundances with longitude, while only the abundance of the latter molecule experiences some longitudinal variation, leading to a slight variation of the planetary flux with phase around 4.3 $\mu$m. In HD~189733b, the homogenization of the chemical composition with longitude is even more marked than in HD~209458b because of the lack of a stratosphere and the rather low eddy coefficient values (compare Figs.~\ref{fig:spaghetti-hd209458b} and \ref{fig:spaghetti-hd189733b}). Because the abundances of H$_2$O, CO, CO$_2$, CH$_4$, NH$_3$, and HCN (the main molecules providing opacity, in order of decreasing importance) vary little between the different sides of HD~189733b at the pressures probed by emission spectra ($>$10$^{-2}$ bar), the impact of the chemical composition on the change of the emission spectrum with phase becomes almost negligible, even around 4.3 $\mu$m because of the reduced longitudinal variation of the abundance of CO$_2$.

\subsection{Transmission spectra of evening and morning limbs}

\begin{figure*}
\centering
\includegraphics[angle=0,width=\textwidth]{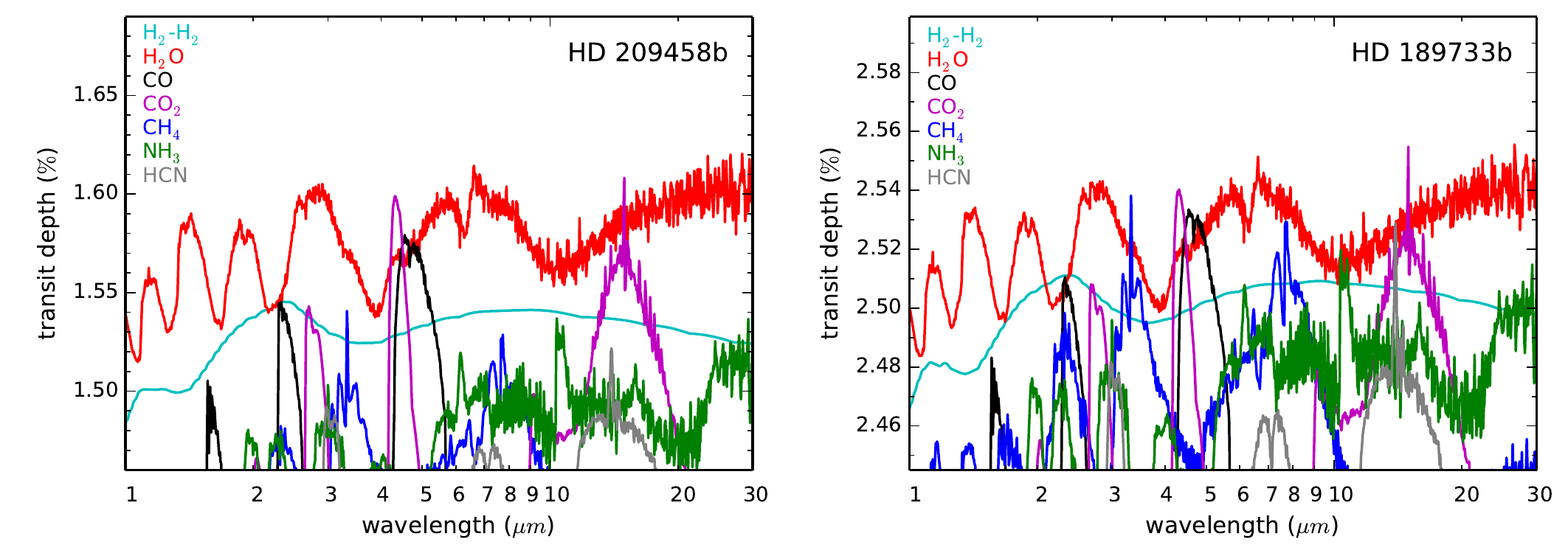}
\caption{Contributions of the different sources of opacity to the transmission spectra of HD~209458b (left panel) and HD~189733b (right panel) smoothed to a resolving power $R$ = 300. The H$_2$-He continuum, whose contribution is similar in shape to the H$_2$-H$_2$ continuum but lower because of the lower abundance of He with respect to H$_2$, is not shown. The spectra have been calculated using the temperature from GCM simulations and the chemical composition from the nominal pseudo two-dimensional chemical model averaged over the whole terminator. Each line represents the transmission spectrum that results from a model in which only the opacity provided by each source is taken into account.} \label{fig:transmission-species}
\end{figure*}

Variations in the composition of the atmosphere between the different sides of the planet may also be probed by transmission spectroscopy. Indeed, it is a priori possible to probe differences in the thermal and chemical structure of the two limbs if observations are able to obtain the transmission spectrum during the first half of the primary transit ingress, which would probe the leading or morning limb, and during the second half of the transit egress, which would probe the trailing or evening limb. Although a non-zero impact parameter during the transit would complicate the situation somewhat and such observations are very challenging today, they may be feasible in the near future. The subject has been addressed theoretically for hot Jupiters such as HD~189733b and HD~209458b in some studies in which the differences between the transmission spectra of leading and trailing limbs are evaluated under different assumptions for the chemical composition of each of the two limbs, either chemical equilibrium or some disequilibrium estimation (\cite{for2010} 2010; \cite{bur2010} 2010). Here we revisit the subject in the light of the molecular abundances calculated in this study with the pseudo two-dimensional chemical model.

To illustrate the possibility that transmission spectroscopy might be able to distinguish between the two limbs of HD~209458b and HD~189733b we show in Fig.~\ref{fig:transmission} the transmission spectrum calculated by adopting the chemical composition and mean temperature of the evening and morning limbs of these two exoplanets. Since we are mainly interested in comparing the spectra at the two limbs and not in comparing with observations, we set the absolute scale of transmission spectra by simply assigning the value of the planetary radius given in section~\ref{sec:model} to the 1 bar pressure level and made no attempt to reproduce the absolute scale of the photometric transit depths derived from observations. In HD~209458b and HD~189733b, the transmission spectrum of the evening limb shows a higher degree of absorption and also the variations of the transit depth with wavelength have a larger amplitude than at the morning limb, whose transmission spectrum is flatter. These differences are mainly due to the different temperature profile of the two limbs. Because the atmosphere at the morning limb is cooler and thus has a smaller scale height than at the evening limb, it becomes more compact, resulting in smaller apparent radii at all wavelengths and a flatter transmission spectrum. In addition to this dependence of the transmission spectrum on temperature, which causes it to shift up or down and to have a more elongated or flattened overall shape, the spectral structure is controlled by the relative abundances of the main species that provide opacity in the atmosphere. Fig.~\ref{fig:transmission-species} shows the relative contributions of the different sources of opacity taken into account in calculating the transmission spectra. Similarly to the emission spectra, in the calculated transmission spectra of HD~209458b and HD~189733b most of the atmospheric opacity at infrared wavelengths is provided by H$_2$O, with CO being important at 2.3 and 4.6 $\mu$m, CO$_2$ at 4.3 and 15 $\mu$m, the H$_2$-H$_2$ continuum at certain wavelengths below 3 $\mu$m, and, in the case of HD~189733b, CH$_4$ having some contribution around 3.3 and 7.7 $\mu$m, NH$_3$ around 10.6 $\mu$m, and HCN at 14 $\mu$m.

Similarly to emission spectra, we evaluated to which extent differences in the chemical composition of evening and morning limbs contribute to the change of the transmission spectrum from one limb to the other. To this purpose, we computed transmission spectra in which we switched the temperature and chemical profiles between the two different limbs. As an example we compare in the insets of left and right panels in Fig.~\ref{fig:transmission} the transmission spectrum of the evening limb with a synthetic spectrum calculated using the temperature structure of the evening limb and the chemical composition of the morning limb. In HD~209458b the two spectra are very similar, except for a different degree of absorption around 4.3 and 15 $\mu$m, which is due to the difference of nearly one order of magnitude in the abundance of CO$_2$ between the two limbs (see Fig.~\ref{fig:abun-hd209458b}). Similarly to emission spectra, the longitudinal homogenization driven by horizontal transport is at the origin of the weak impact of other molecules on the variation of transmission spectra between both limbs. Because the abundances of CO and H$_2$O are very similar in both limbs, and other molecules such as CH$_4$, NH$_3$, and HCN contribute little to the atmospheric opacity at infrared wavelengths because of their rather low abundances, the only chemical effect contributing to the change of the transmission spectrum from one limb to the other of HD~209458b is restricted to carbon dioxide. In HD~189733b, the even stronger longitudinal homogenization of abundances (compare Figs.~\ref{fig:spaghetti-hd209458b} and \ref{fig:spaghetti-hd189733b}) diminishes the extent of chemical effects, which are now restricted to a very weak change of the absorption around 4.3 and 15 $\mu$m (see inset in HD~189733b's panel of Fig.~\ref{fig:transmission}), again due to a slight increase in the abundance of CO$_2$ when moving from the evening limb to the morning one (see Fig.~\ref{fig:abun-hd189733b}).

\section{Summary}

We have developed a pseudo two-dimensional model of a planetary atmosphere that takes into account thermochemical kinetics, photochemistry, vertical mixing, and horizontal transport, and allows one to calculate the distribution with altitude and longitude of the main atmospheric constituents. Horizontal transport was modeled through a uniform zonal wind and thus the model is best suited for studying the atmosphere of planets whose circulation dynamics is dominated by an equatorial superrotating jet, as is expected to be the case of hot Jupiters. We therefore applied the model to study the atmospheres of the well-known exoplanets HD~209458b and HD~189733b. We used the temperature structure from GCM simulations and parameterized the turbulent mixing in the vertical direction using an eddy coefficient profile, which was calculated by following the behavior of passive tracers in GCM simulations, a method that results in substantially lower eddy values, by a factor of 10-100 in HD~209458b and of 10-1000 in HD~189733b, than previous estimates based on cruder methods.

\emph{Molecular abundances homogenized with longitude to values typical of the hottest dayside regions. --} We found that the distribution of molecules in the atmospheres of HD~209458b and HD~189733b is quite complex because of the interplay of the various (photo)chemical and dynamical processes at work, which form, destroy, and transport molecules throughout the atmosphere. Much of the distribution of the atmospheric constituents is driven by the strong zonal wind, which reaches speeds of a few km s$^{-1}$, and the limited extent of vertical transport, with relatively low eddy diffusion coefficients below 10$^9$ cm$^2$ s$^{-1}$ around the 1 bar pressure level, resulting in an important homogenization of molecular abundances with longitude, in particular in the atmosphere of HD~189733b, which lacks a stratosphere and has quite low eddy diffusion coefficients. Moreover, molecular abundances are quenched horizontally to values typical of the hottest dayside regions, and therefore the composition of the cooler nightside regions is highly contaminated by that of warmer dayside regions. In hot Jupiters with a temperature inversion, such as HD~209458b, the longitudinal homogenization of molecular abundances is not as marked as in planets lacking a stratosphere, such as HD~189733b. In general, the cooler the planet, the stronger the homogenization of the chemical composition with longitude. Furthermore, in cooler planets such as hot Neptunes orbiting M dwarfs (e.g., GJ 436b) the temperature contrast between day and nightsides decreases because the cooling rate scales with the cube of temperature (e.g., \cite{lew2010} 2010), and therefore the composition is expected to be even more homogeneous with longitude than in warmer planets such as HD~209458b and HD~189733b. However, unlike hot Jupiters, hot Neptunes may have an atmospheric metallicity much higher than solar (\cite{lin2011} 2011; \cite{mos2013} 2013; \cite{agu2014} 2014; \cite{ven2014} 2014), which makes it interesting to investigate the extent of the spatial variation of molecular abundances in their atmospheres.

\emph{Low methane content. --} A major consequence of our pseudo two-dimensional chemical model is that methane reaches quite low abundances in the atmospheres of HD~209458b and HD~189733b, lower than the values calculated by previous one-dimensional models. The main reason for the low CH$_4$ abundance is that most of the atmosphere is contaminated by the hottest dayside regions, where the chemical equilibrium abundance of CH$_4$ is the lowest. The calculated mixing ratio of CH$_4$ in the dayside of HD~209458b is significantly below the values inferred from observations, which points to some fundamental problem in either the chemical model or the observational side. If the strength of vertical transport is substantially higher than in our nominal model, the calculated abundance of some molecules such as CH$_4$ and NH$_3$ would experience significant enhancement, especially in HD~189733b, although a conflict with observations would still exist regarding CH$_4$ in the dayside of HD~209458b.

\emph{Variability of planetary spectra driven by thermal, rather than chemical, gradients. --} An important consequence of the strong longitudinal homogenization of molecular abundances in the atmospheres of HD~209458b and HD~189733b is that the variability of the chemical composition has little effect on the way the emission spectrum is modified with phase and on the changes of the transmission spectrum from the transit ingress to the egress. Temperature variations and not chemical gradients are therefore at the origin of these types of variations in the planetary spectra. Only the longitudinal variation of the abundance of CO$_2$, of nearly one order of magnitude, in the atmosphere of HD~209458b, is predicted to induce variations in the planetary spectra around 4.3 and 15 $\mu$m. We note, however, that an inhomogenous distribution of clouds and/or hazes (none of them included in our model) may induce important variations in the emission spectra with phase and in the transmission spectra from one limb to the other. These variations are best characterized at short wavelengths. Indeed, there is evidence of the presence of hazes in the atmosphere of HD~189733b (\cite{lec2008} 2008; \cite{sin2009} 2009), and an inhomogeneous distribution of clouds has recently been inferred for the hot Jupiter Kepler 7b (\cite{dem2013} 2013).

The main drawback of our pseudo two-dimensional chemical model is the oversimplification of atmospheric dynamics, which is probably adequate for equatorial regions, but not at high latitudes. Ideally, GCM simulations coupled to a robust chemical network would provide an even more realistic view of the distribution of molecules in the atmospheres of HD~209458b and HD~189733b, but such calculations are very challenging from a computational point of view. Telescope facilities planned for the near or more distant future, such as the James Webb Space Telescope, Spica, and EChO, will be able to test some of the predictions of our pseudo two-dimensional model, in particular the low abundance of methane in the two planets and the important longitudinal homogenization of the chemical composition.

\begin{acknowledgements}

We thank our anonymous referee for insightful comments which helped to improve this article. We acknowledge Adam P. Showman and Jonathan J. Fortney for the use of the SPARC/MITgcm code, Vincent Hue for useful discussions on photochemical models, Sergio Blanco-Cuaresma and Christophe Cossou for their help with Python and Fortran, and Vincent Eymet and Philip von Paris for kindly helping to validate the line--by--line radiative transfer code. M. A. and F. S. acknowledge support from the European Research Council (ERC Grant 209622: E$_3$ARTHs). O.V. acknowledges support from the KU Leuven IDO project IDO/10/2013 and from the FWO Postdoctoral Fellowship Program. Computer time for this study was provided by the computing facilities MCIA (M\'esocentre de Calcul Intensif Aquitain) of the Universit\'e de Bordeaux and of the Universit\'e de Pau et des Pays de l'Adour.

\end{acknowledgements}

\begin{appendix}

\section{Calculation of planetary spectra} \label{app:spectra}

To investigate the influence of the physical and chemical structure of the atmospheres of HD~209458b and HD~189733b on their transmission and emission spectra, we developed a line-by-line radiative transfer code that is independent of the pseudo two-dimensional chemical code. Currently, the code is limited because it is one-dimensional in the sense that the atmosphere is divided into various layers in the vertical direction (typically 60 spanning the 10-10$^{-6}$ bar pressure range) and each layer is assumed to be homogeneous with longitude and latitude. Therefore, each layer is characterized by a given pressure, temperature, and chemical composition, and longitudinal and latitudinal gradients are neglected. For transmission spectra, the physical and chemical profile in the vertical direction at either the east or west limb, or a mean of the profiles at both limbs, can be used. For emission spectra, the limitations caused by the one-dimensional character of the code can be partially alleviated by adopting thermal and chemical vertical profiles averaged in some manner over the hemisphere facing the observer.

It is common in infrared spectroscopy to use the wavenumber with units of cm$^{-1}$, instead of frequency or wavelength, and we therefore adopt this choice hereafter as well. At this stage, there are various sources of opacity included in the code. On the one hand, we consider collision induced absorption (CIA) by H$_2$-H$_2$, for which available absorption coefficients cover the wavelength range 10-25,000 cm$^{-1}$ and temperatures between 60 and 7000 K (\cite{bor2001} 2001; \cite{bor2002} 2002), and by H$_2$-He, in which case absorption coefficients in the wavelength range 10-25,000 cm$^{-1}$ and for temperatures in the range 100-7000 K are available (\cite{bor1989} 1989, 1997; \cite{borfro1989} 1989). CIA absorption coefficients scale with the square of pressure and thus become the dominant source of opacity at high pressures, usually above 1 bar. On the other hand, we consider spectroscopic transitions (mostly ro-vibrational transitions lying at infrared wavelengths) of H$_2$O, CO, and CO$_2$, whose data are taken from HITEMP (\cite{rot2010} 2010), and of CH$_4$, NH$_3$, and HCN, for which data from HITRAN (\cite{rot2009} 2009) are adopted.

The spectral region of interest is divided into a certain number of spectral bins, whose widths are determined by the spectral resolution imposed. In each layer of the atmosphere, the contribution of a spectroscopic transition $j$ (centered on a wavenumber $\tilde{\nu_j}$ and which belongs to a species $i$) to the absorption coefficient $k(\tilde{\nu}_l)$ in a spectral bin $l$ (centered on a wavenumber $\tilde{\nu}_l$ and having a width $\Delta \tilde{\nu}_l$), which we may label as $k_{ij}^S(\tilde{\nu}_l)$, can be expressed as
\begin{equation}
k_{ij}^S(\tilde{\nu}_l) = S_j(T) n_i \int_{\tilde{\nu}_l-\Delta \tilde{\nu}_l / 2}^{\tilde{\nu}_l+\Delta \tilde{\nu}_l / 2} \phi_j (\tilde{\nu}' - \tilde{\nu_j}) d \tilde{\nu}' \frac{1}{\Delta \tilde{\nu}_l}, \label{eq:kabs}
\end{equation}
where $S_j(T)$ is the line intensity of the spectroscopic transition $j$, which depends on the temperature $T$ and is usually given with units of cm$^{-1}$/(molecule cm$^{-2}$) in the HITRAN and HITEMP databases, $n_i$ is the number density of species $i$ in the atmospheric layer where the absorption coefficient is to be evaluated, and $\phi_j$ is the line profile function of transition $j$, which has units of inverse of wavenumber (i.e., cm) and must be normalized such that the integral of $\phi_j(\tilde{\nu}' - \tilde{\nu_j})$ from $\tilde{\nu}' - \tilde{\nu_j}=-\infty$ to $\tilde{\nu}' - \tilde{\nu_j}=+\infty$ yields unity. The absorption coefficient $k(\tilde{\nu}_l)$ has units of cm$^{-1}$. The integral in Eq.~(\ref{eq:kabs}) extends between the lower and upper wavenumber edges of the spectral bin $l$. The line profile function $\phi_j(\tilde{\nu}' - \tilde{\nu_j})$ is taken as a Voigt profile, which results from the convolution of a Gaussian and a Lorentzian profile, and thus accounts for the Doppler and pressure broadening of spectral lines in each layer of the atmosphere due to thermal motions and collisions, respectively. The Voigt profile function is calculated numerically with a routine based on an implementation of \cite{hum1982}'s algorithm by \cite{kun1997} (1997).

Ideally, extremely high spectral resolution would be desirable to properly resolve the narrowest line profiles, although in practice this is too expensive in terms of computing time. For the calculations carried out here we adopted a spectral resolution of 0.03 cm$^{-1}$, which is on the order of the line widths in the layers that provide most atmospheric opacity and has been found to be high enough to yield relative errors below 1 \% in the computed spectra. Spectroscopic transitions lying farther away than 50 cm$^{-1}$ of a given spectral bin $l$ have not been taken into account when computing the absorption coefficient $k(\tilde{\nu}_l)$. In addition, a cutoff in the line intensity $S_j$(296 K) of 10$^{-40}$ cm$^{-1}$/(molecule cm$^{-2}$) was adopted to neglect weak lines and speed up the calculations. This is perhaps the most delicate aspect because weak lines are numerous, especially in the HITEMP line lists, and at high temperatures the sum of all them results in non-negligible opacity enhancements in certain spectral regions. Calculations carried out with different line intensity cutoffs in selected spectral regions indicate that the relative error in the calculated spectra is at most 10 \% in the hottest case studied here (the dayside emission spectrum of HD~209458b), and lower than 5 \% in the remaining computed spectra. Currently, the code does not take into account light scattering, which becomes important at wavelengths shorter than $\sim$1 $\mu$m, and therefore calculated spectra are reliable at infrared wavelengths, but not in the visible region of the electromagnetic spectrum. The code does not take into account the Doppler shift of spectral lines due to atmospheric winds either, an effect that may be observable in high-resolution spectra of hot Jupiters, where winds are strong (\cite{sne2010} 2010; \cite{mil2012} 2012; \cite{sho2013} 2013).

At each atmospheric layer, the absorption coefficient $k(\tilde{\nu}_l)$ in each wavenumber bin $l$ is calculated as the sum of the contributions from all the spectroscopic transitions of the various absorbing species included, together with the contributions of the CIA couples, that is, 
\begin{equation}
k(\tilde{\nu}_l) = \sum_i \sum_j k_{ij}^S (\tilde{\nu}_l) + \sum_m k_m^{CIA} (\tilde{\nu}_l), \label{eq:kabs-add}
\end{equation}
where the sum in $m$ extends to the H$_2$-H$_2$ and H$_2$-He couples.

After calculating the absorption coefficient in each wavenumber bin and at each atmospheric layer, computing the transmission and emission planetary spectra becomes straightforward, provided scattering is not considered. To calculate the transmission spectrum, the optical depth $\tau(\tilde{\nu}_l,b)$ along a tangential line of sight intersecting the planet's atmosphere is computed as a function of the impact parameter $b$ in each spectral interval $l$ as
\begin{equation}
\tau(\tilde{\nu}_l,b) = \sum_{h=1}^{N(b)} k_h(\tilde{\nu}_l) \Delta \ell_h(b), \label{eq:transmission}
\end{equation}
where $\Delta \ell_h(b)$ is the path length (in cm) intersected by layer $h$ along the tangential line of sight at impact parameter $b$, and the sum in $h$ extends to all atmospheric layers $N(b)$ intersected by the tangential line of sight. The apparent radius of the planet in each wavenumber interval is then retrieved as the impact parameter for which the optical depth along the tangential line of sight becomes 2/3. This latter value is rather arbitrary and is mainly chosen for similarity with the definition of a stellar photosphere, although it is very close to the value of 0.56 inferred by \cite{lec2008} (2008) and is not critical to derive the apparent radius of the planet. To obtain the emission spectrum, the emergent specific intensity along the observer's line of sight is computed as a function of impact parameter in each wavenumber bin. We thus need to solve the equation of radiative transfer along the various paths pointing toward the observer that pass through the planetary atmosphere at different impact parameters. As long as the different atmospheric layers are homogeneous, the equation of radiative transfer can be solved sequentially from the back to the front for each of the atmospheric layers intersected by the path. For each intersected layer, the equation of radiative transfer reads (see e.g. \cite{ryb2004} 2004)
\begin{equation}
I_{\tilde{\nu}} = e^{-\tau_{\tilde{\nu}}} \big[ I_{\tilde{\nu}}^0 + B_{\tilde{\nu}}(T) \big(e^{\tau_{\tilde{\nu}}} - 1\big) \big], \label{eq:emission}
\end{equation}
where the subscript $\tilde{\nu}$ indicates wavenumber dependence, $I_{\tilde{\nu}}^0$ and $I_{\tilde{\nu}}$ are the incoming and outgoing specific intensities that enter and emerge, respectively, from the current layer along a given path, $\tau_{\tilde{\nu}}$ is the optical depth along the path within the current layer, and $B_{\tilde{\nu}}(T)$ is Planck's function. The final emission spectrum is then computed by averaging the wavenumber-dependent specific intensity, calculated as a function of the impact parameter, over the projected area of the emitting hemisphere.

The line-by-line radiative transfer code was checked against the suite of radiative transfer tools 'kspectrum'\footnote{See \texttt{http://code.google.com/p/kspectrum/}}, which has been widely used to model the atmosphere of solar system planets such as Venus (\cite{eym2009} 2009).

\end{appendix}

\end{document}